%% file: Ly2015_OIII4363.arxiv.v3.tex
\documentclass[numberedappendix]{emulateapj}
\usepackage{amsmath,amssymb,graphics}
\usepackage[colorlinks=true,  citecolor=blue,  
  linkcolor=blue,  menucolor=blue, urlcolor=blue,
  linkbordercolor={0 0 1}, frenchlinks=True, breaklinks]{hyperref}
\usepackage{multirow}
\begin{document}
\newcommand{\Sname}{$\mathcal{MACT}$}
\newcommand{\Ha}{H$\alpha$}
\newcommand{\Hb}{H$\beta$}
\newcommand{\Hg}{H$\gamma$}
\newcommand{\Hd}{H$\delta$}
\newcommand{\Hyd}{{\rm H}}
\newcommand{\Hae}{\Hyd\alpha}
\newcommand{\Hbe}{\Hyd\beta}
\newcommand{\Hge}{\Hyd\gamma}
\newcommand{\Hde}{\Hyd\delta}
\newcommand{\Lya}{Ly$\alpha$}
\newcommand{\NII}{[{\rm N}\,\textsc{ii}]}
\newcommand{\OIII}{[{\rm O}\,\textsc{iii}]}
\newcommand{\OII}{[{\rm O}\,\textsc{ii}]}
\newcommand{\SII}{[{\rm S}\,\textsc{ii}]}
\newcommand{\NeIII}{[{\rm Ne}\,\textsc{iii}]}
\newcommand{\OIIIa}{\OIII\,$\lambda$4363}
\newcommand{\OI}{[{\rm O}\,\textsc{i}]}
\newcommand{\OIl}{\OI\,$\lambda$6300}

\newcommand{\OIIIA}{[\textsc{O iii}]{\rm-A}}
\newcommand{\zphotf}{z_{\rm phot}}
\newcommand{\zspecf}{z_{\rm spec}}
\newcommand{\zphot}{$\zphotf$}
\newcommand{\zspec}{$\zspecf$}
\newcommand{\Rcf}{R_{\rm C}}
\newcommand{\Rc}{$\Rcf$}

\newcommand{\Pagel}{$R_{23}$}
\newcommand{\Oratio}{$O_{32}$}
\newcommand{\Te}{$T_e$}
\newcommand{\OH}{12\,+\,$\log({\rm O/H})$}
\newcommand{\OHm}{12\,+\,\log({\rm O/H})}
\newcommand{\zsun}{$Z_{\sun}$}
\newcommand{\MB}{$M_B$}
\newcommand{\fifth}{$Z$=0.004}
\newcommand{\solar}{$Z$=0.02}

\newcommand{\rNB}{{\rm NB}}
\newcommand{\nOII}{$\sim$1,300}

\newcommand{\HaHbi}{2.86}

\newcommand{\EBV}{E(B-V)}
\newcommand{\EBVa}{$E$($B$--$V$)}
\newcommand{\cc}{cm$^{-3}$}
\newcommand{\mm}{$\mu$m}
\newcommand{\Msun}{$M_{\sun}$}
\newcommand{\Mstar}{$M_{\star}$}
\newcommand{\iyr}{yr$^{-1}$}

\newcommand{\MZ}{\Mstar--$Z$}
\newcommand{\MZR}{MZR}
\newcommand{\zmin}{0.05}

\newcommand{\za}{0.01}
\newcommand{\zb}{1.62}
\newcommand{\areaa}{870.4}
\newcommand{\areab}{788.7}

\newcommand{\Nem}{9264}
\newcommand{\Nspec}{3243}
\newcommand{\Nspecg}{1911}
\newcommand{\Nspecgz}{1493}
\newcommand{\NMMTspec}{1820}
\newcommand{\NMMTspecg}{845}
\newcommand{\NKeckspec}{1423}
\newcommand{\NKeckspecg}{1313}
\newcommand{\Ndet}{66}
\newcommand{\Nrel}{98}
\newcommand{\Ntot}{164}
\newcommand{\Ndetf}{19}
\newcommand{\NMMTf}{37}
\newcommand{\NKeckf}{33}

\newcommand{\NOIIIMMT}{67}
\newcommand{\NOIIIKeck}{119}

\newcommand{\NHa}{33}
\newcommand{\NXMPG}{16}
\newcommand{\zslope}{-2.32^{+0.52}_{-0.26}}
\newcommand{\DsSFR}{$\Delta({\rm sSFR})_{\rm MS}$}

\newcommand{\GALEX}{{\it GALEX}}
\newcommand{\FUV}{{\it FUV}}
\newcommand{\NUV}{{\it NUV}}

\newcommand{\TM}{\tablenotemark{M}}
\newcommand{\TA}{\tablenotemark{a}}
\newcommand{\TB}{\tablenotemark{b}}
\newcommand{\TC}{\tablenotemark{c}}
\newcommand{\TD}{\tablenotemark{d}}

\newcommand{\pa}{\phantom{1}}

\newcommand{\SFRA}{--0.19$\pm$0.79}
\newcommand{\SFRM}{--0.18}
\newcommand{\sSFRA}{10$^{-8.4}$}
\newcommand{\sSFRt}{240 Myr}
\newcommand{\MassA}{$1.5\times10^8$}
\newcommand{\MassM}{$1.3\times10^8$}
\newcommand{\AgeA}{8.0}

\newcommand{\SMACT}{2}
\newcommand{\SSample}{3}
\newcommand{\SProp}{4}
\newcommand{\SSED}{4.4}
\newcommand{\TSFR}{15}
\newcommand{\TSFRND}{16}
\newcommand{\FLum}{24}

\title{The Metal Abundances across Cosmic Time ($\mathcal{MACT}$) Survey. II.
  Evolution of the Mass--Metallicity Relation over 8 Billion Years, Using
  \OIII$\lambda$4363 \AA\ Based Metallicities}

\author{Chun Ly,\altaffilmark{1} Matthew A. Malkan,\altaffilmark{2}
  Jane R. Rigby,\altaffilmark{1} and Tohru Nagao\altaffilmark{3}}

\submitted{Received 2016 February 1; revised 2016 June 14; accepted 2016 June 19;
  published 2016 September 1}
\shorttitle{Evolution of the Mass-Metallicity Relation}
\shortauthors{Ly et al.}
\email{astro.chun@gmail.com}

\altaffiltext{1}{Observational Cosmology Laboratory, NASA Goddard Space Flight Center,
  8800 Greenbelt Road, Greenbelt, MD 20771, USA}
\altaffiltext{2}{Department of Physics and Astronomy, UCLA, Los Angeles, CA 90095, USA}
\altaffiltext{3}{Research Center for Space and Cosmic Evolution, Ehime University,
    Matsuyama 790-8577, Japan}

\begin{abstract}
  We present the first results from MMT and Keck spectroscopy for a large sample of
  $0.1 \leq z \leq 1$ emission-line galaxies selected from our narrow-band imaging in
  the Subaru Deep Field. We measured the weak \OIIIa\ emission line for \Ntot\
  galaxies (\Ndet\ with at least 3$\sigma$ detections, and \Nrel\ with significant
  upper limits). The strength of this line is set by the electron temperature for the
  ionized gas. Because the gas temperature is regulated by the metal content, the
  gas-phase oxygen abundance is inversely correlated with \OIIIa\ line strength.
  Our temperature-based metallicity study is the first to span $\approx$8 Gyr of
  cosmic time and $\approx$3 dex in stellar mass for low-mass galaxies,
  $\log{\left(M_{\star}/M_{\sun}\right)}\approx6.0$--9.0. Using extensive
  multi-wavelength photometry, we measure the evolution of the stellar mass--gas
  metallicity relation and its dependence on dust-corrected star formation rate (SFR).
  The latter is obtained from high signal-to-noise Balmer emission-line measurements.
  Our mass-metallicity relation is consistent with Andrews \& Martini at $z\leq0.3$,
  and evolves toward lower abundances at a given stellar mass,
  $\log{({\rm O/H})}\propto(1+z)^{\zslope}$. We find that galaxies with lower
  metallicities have higher SFRs at a given stellar mass and redshift, although the
  scatter is large ($\approx$0.3 dex) and the trend is weaker than seen in local
  studies. We also compare our mass--metallicity relation against predictions from
  high-resolution galaxy formation simulations, and find good agreement with models
  that adopt energy- and momentum-driven stellar feedback. We have identified \NXMPG\
  extremely metal-poor galaxies with abundances less than a tenth of solar; our most
  metal-poor galaxy at $z\approx0.84$ is similar to I Zw 18.
\end{abstract}


\defcitealias{ly07}{Ly07}
\defcitealias{MACTI}{Paper I}
\defcitealias{and13}{AM13}


\section{INTRODUCTION}
\label{1}

The chemical enrichment of galaxies, driven by star formation and modulated by gas
flows from supernova and cosmic accretion, is key for understanding galaxy formation
and evolution. The primary method for measuring metal abundances is spectroscopy of
nebular emission lines. The strongest lines can be observed in the optical and
near-infrared at $z\lesssim3$ from the ground and space.

The most reliable metallicity measurements are based on the flux ratio of the \OIIIa\
line against \OIII$\,\lambda$5007. The technique is called the \Te\ method, because
it determines the electron temperature (\Te) of the gas, and hence the gas-phase
oxygen-to-hydrogen (O/H) abundance \citep{all84,izo06b}.
However, detecting \OIIIa\ is difficult, because it is weak and almost undetectable in
metal-rich galaxies. For example, only 0.3\% of the strongly star-forming galaxies
in the Sloan Digital Sky Survey (SDSS) have $2\sigma$ or better detections of \OIIIa\
\citep{izo06b,nag06}.

After enormous observational efforts to increase the number of galaxies with \Te-based
metallicities in the local universe \citep[e.g.,][]{bro08,berg12,izo12}, and at
$z\gtrsim0.2$ \citep{hoy05,kak07,hu09,atek11,amo15,amo14,ly14,ly15}, the total sample
size of $\ge3\sigma$ \OIIIa\ detections is 174 galaxies.

\Te-based metallicities are even harder to measure at $z\gtrsim0.2$. Thus the evolution
of the stellar mass--gas metallicity (\MZ) relation, and its dependence on star
formation rate (SFR), has only been studied using empirical or theoretical
estimates based on strong nebular emission lines
\citep[e.g., \NII$\,\lambda$6583, \OIII, \OII, \Ha, \Hb;][]{pag79,pet04}, which have to
be calibrated against \Te-based metallicities in local galaxies and \ion{H}{2}
regions \citep[e.g.,][]{KK04,erb06,mai08,hai09,hay09,lam09,man09,man10,thu10,mou11,
  rig11,Wel11,zah11,zah12,zah13,zah14,hunt12,nak12,xia12,yab12,yat12,bel13,gua13,hen13a,
  hen13b,mom13,pir13,cul14,ly14,ly15,mai14,sal14,tro14,whi14b,yab14,
  yab15,rey15,wuy12,hay15,san15}.

However, there are problems with these ``strong-line'' metallicity calibrations. For
example, depending on which one is used,  the shape and normalization of the \MZ\
relation differ significantly at $\sim$1 dex
\citep[see Figure 2 in][]{kew08}.\footnote{We note that while the \Te\ method is
  affected by properties of the ionized gas \citep[e.g., optical depth, density,
    ionization parameter, non-equilibrium electron energy,
    temperature fluctuation;][]{est99,hag06,nic14}, most of these effects
  also apply to strong-line diagnostics \citep{nic14}. Thus, while the \Te\ method
  is less reliable than was initially thought \citep{sea54}, measuring the
  electron temperature currently remains the preferred way to determine gas
  metallicities.}
Therefore, studies cannot examine the evolution of the \MZ\ relation unless they use
the same metallicity calibration for all galaxies. This method of comparing
metallicities on a relative level is only valid if the physical conditions of the
interstellar gas (e.g., gas density, ionization, N/O abundance) do not evolve.
However, clear evidence now suggests that the physical conditions of the gas in
high-$z$ galaxies are significantly different from those in local galaxies. For
example, $z\gtrsim1$ star-forming galaxies are known to be offset on the
Baldwin--Phillips--Terlevich (``BPT'') diagnostic diagrams
\citep[\OIII\,$\lambda$5007/\Hb\ vs. \NII\,$\lambda$6583/\Ha;][]{bal81}
from local star-forming galaxies
\citep[e.g.,][and references therein]{sha05,liu08,fin09,hai09,bian10,
  rig11,kew13b,ste14,sha15}.
This offset is seen as a higher \OIII/\Hb\ ratio at fixed \NII\,$\lambda$6583/\Ha.
It has been tentatively attributed to a higher ionization parameter, harder ionizing
spectrum, and/or higher electron density in star-forming regions at higher redshifts
\citep[e.g.,][]{bri08,kew13a}.
Alternatively, recent studies of strongly star-forming galaxies at $z\approx0.1$--0.35
and $z\sim2$ indicate they have enhanced N/O abundance ratios compared to typical
galaxies at $z\sim0.1$ from SDSS, resulting in stronger \NII\,$\lambda$6583 line
emission for given strengths of the oxygen forbidden lines (e.g.,
\citealt{amo10,mas14}). Depending on the explanation for the higher N/O, results
involving commonly used metallicity estimates from the \NII/\Ha\ ratio \citep{pet04}
will overestimate oxygen abundances by $\approx$0.25--1 dex.

\subsection{Sample Selection}
To address the lack of \OIIIa\ measurements at higher redshifts, and outstanding
issues with gas metallicity calibrations for higher redshift galaxies, we conducted
a spectroscopic survey called ``Metal Abundances across Cosmic Time''
\citep[\Sname;][hereafter Paper I]{MACTI} to obtain deep (2--12 hr) rest-frame
optical spectra of $z\lesssim1$ star-forming galaxies with Keck and MMT. The primary
goal of the survey was to obtain reliable measurements of the gas-phase metallicity
and other physical properties of the interstellar medium (ISM) in galaxies, such as
the SFR, gas density, ionization parameter, dust content, and the source of
photoionizing radiation (star formation and/or active galactic nucleus, AGN).
\Sname\ is unique among previous spectroscopic surveys because it is the first
to use the \Te\ method to measure the evolution of the \MZ\ relation over
$\approx$8 billion years. In addition, the galaxy sample of \Sname\ encompasses
nearly 3 dex in stellar mass, including dwarfs as low as
$M_{\star}\sim3\times10^6$ \Msun\ and $3\times10^7$ \Msun\ at $z\sim0.1$ and $z\sim1$,
respectively.
The \Sname\ survey targeted $\approx$1900 galaxies in the Subaru Deep Field
\citep[SDF;][]{kas04} that have excess flux in narrow-band and/or intermediate-band
filters, which is now understood to be produced by nebular emission lines from star
formation or AGNs \citep[e.g.,][and references therein]{ly07,newha}.

In this paper, Paper II, we focus on the first results from \Ndet\ galaxies with at
least S/N = 3\footnote{Of the \Ndet\ \OIIIa-detected galaxies, 31 have detections
  above S/N = 5.} detections of \OIIIa\ at $z=\zmin$--0.95 (average of
$z=0.53\pm0.25$; median of 0.48), and robust \OIIIa\ upper limits for \Nrel\ galaxies
at $z=0.04$--0.96 (average of $z=0.52\pm0.23$; median of 0.48). We refer to the
collective of these galaxies as the ``\OIIIa-detected and \OIIIa-non-detected
samples.'' For the \OIIIa-non-detected galaxies, we require an \OIII$\lambda$5007
detection that is at S/N $\gtrsim$ 100 and S/N $<$ 3 for \OIIIa. This work expands
on our previous sample of spectroscopic detections of \OIIIa\ \citep{ly14} by more
than threefold. In a forthcoming paper, we will use our sample with \OIIIa\
measurements to recalibrate the strong-line metallicity diagnostics for these
galaxies at $z\approx0.5$.

We refer readers to \citetalias{MACTI} for more details on the \Sname\ survey and
our primary sample for Paper II. Specifically, Section \SMACT\ in \citetalias{MACTI}
describes the full galaxy sample and optical spectroscopy, Section \SSample\ in
\citetalias{MACTI} describes the \OIIIa-detected and \OIIIa-non-detected sample
selection, and Section \SProp\ in \citetalias{MACTI} describes the interstellar
(i.e., \Te-based metallicity, dust attenuation) and stellar properties (i.e., SFR,
stellar mass) of \OIIIa-detected and \OIIIa-non-detected galaxies.
The outline of this Paper II is as follows.

In Section~\ref{sec:AGN}, we discuss the identification of a small number of AGNs
or low-ionization nuclear emitting regions \citep[LINERs;][]{hec80} that contaminate
our galaxy sample.
In Section~\ref{sec:Results}, we present our five main results: (1) a large sample of
extremely metal-poor galaxies at $z\gtrsim0.1$, (2) comparison of our samples against
other star-forming galaxies on the \Mstar--SFR projection, (3) the similarity of
these metal-poor galaxies to typical star-forming galaxies at high-$z$, (4) the
evolution of the \Te-based \MZ\ relation, and (5) the secondary dependence of the
\MZ\ relation on SFR.
In Section~\ref{sec:Disc}, we compare our \MZ\ relation against predictions from
theoretical and numerical simulations, discuss the selection function of our survey,
and compare our survey to previous \Te-based studies. We summarize results
in Section~\ref{sec:End}.

Throughout this paper, we adopt a flat cosmology with $\Omega_{\Lambda}=0.7$,
$\Omega_M=0.3$, and $H_0=70$ km s$^{-1}$ Mpc$^{-1}$. Magnitudes are reported on the AB
system \citep{oke74}. For reference, we adopt \OH$_{\sun}$ = 8.69 \citep{pri01}
as solar metallicity, \zsun.
Unless otherwise indicated, we report 68\% confidence measurement uncertainties, and
``\OIII'' alone refers to the 5007 \AA\ emission line.


\section{Contamination from LINERs and AGNs}
\label{sec:AGN}


\begin{figure*}
  \epsscale{1.1}
  \plotone{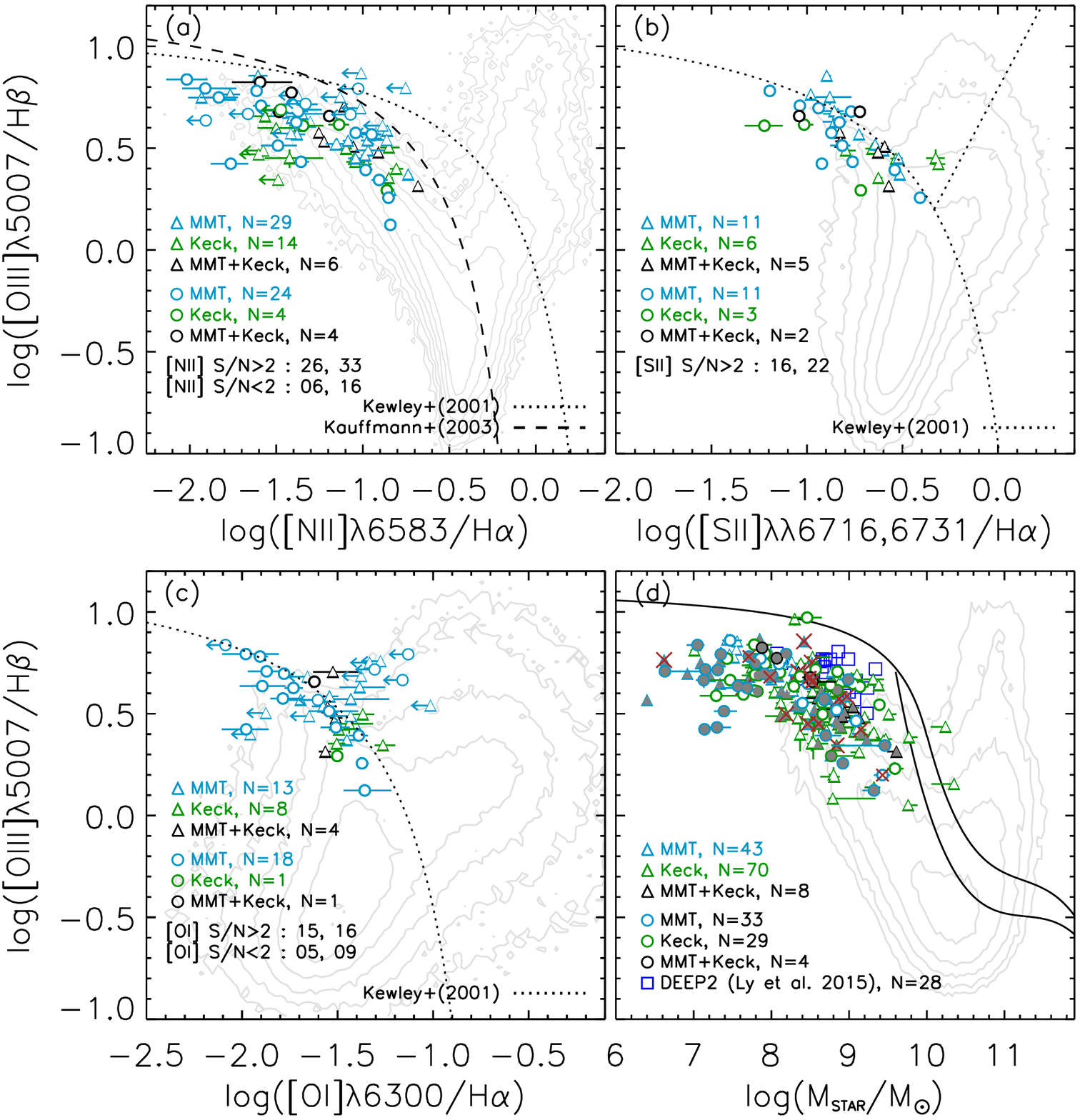}
  \caption{BPT line-ratio diagnostics \citep{bal81,vei87} and the MEx diagram
    \citep{jun14} to distinguish and exclude AGNs and LINERs for our \OIIIa-detected
    (circles) and \OIIIa-non-detected (triangles) samples. The $x$-axes show
    $\log{\left(\NII\,\lambda6583/\Hae\right)}$ (a),
    $\log{\left(\SII\,\lambda\lambda6716,\,6731/\Hae\right)}$ (b),
    $\log{\left(\OI\,\lambda6300/\Hae\right)}$ (c), and
    $\log{\left(M_{\star}/M_{\sun}\right)}$ (d; see Section~\SSED\ of \citetalias{MACTI}),
    while the $y$-axes show $\log{\left(\OIII/\Hbe\right)}$. The MMT, Keck, and the
    MMT+Keck samples are shown in light blue, green, and black, respectively. Upper
    limits (left arrows) on \NII\ and \OI\ fluxes are provided at 2$\sigma$
    confidence. For panel (d), gray-filled circles and triangles indicate SDF galaxies
    that have \NII\ measurements. The \cite{ly15} $z\sim0.8$ DEEP2 \OIIIa\ sample is
    shown as dark blue squares in (d). Dotted lines show the \cite{kew01} criteria
    that separate AGNs from star-forming galaxies (Equations
    (\ref{eqn:BPT1})--(\ref{eqn:BPT3})). The \cite{kau03} criterion is also shown in
    panel (a) as the dashed line. AGNs and LINERs are indicated by brown crosses
    in panel (d).}
  \label{fig:BPT_MEx}
\end{figure*}

A possible concern is whether any of the \OIIIa-detected and \OIIIa-non-detected
galaxies harbor LINERs, or the narrow-line regions of Seyfert nuclei.  When either
of these are present, the gas may not be entirely ionized by young stars.
A strong \OI$\,\lambda$6300 emission line is a defining characteristic of LINERs,
while high \OIII\,$\lambda$5007/\Hb, \NII\,$\lambda$6583/\Ha, and
\SII\,$\lambda\lambda$6716, 6731/\Ha\ ratios indicate a Seyfert 2 AGN.
We classify each of our galaxies by their location on the three standard BPT diagrams
\citep{bal81,vei87}. These are illustrated in Figure~\ref{fig:BPT_MEx}.
For our \OIIIa-detected sample, 32, 16, and 20 galaxies have measurements of \NII/\Ha,
\SII/\Ha, and \OI/\Ha, respectively. These line ratios are also available for 49, 22,
and 25 galaxies from the \OIIIa-non-detected sample, respectively.

We define AGNs as those that meet the \cite{kew01} criteria:
\begin{eqnarray}
  y \geq \frac{0.61}{x_1 - 0.47} + 1.19,\label{eqn:BPT1}\\
  y \geq \frac{0.72}{x_2 - 0.32} + 1.30,\label{eqn:BPT2}\\
  y \geq \frac{0.73}{x_3 + 0.59} + 1.33,{\rm~where}\label{eqn:BPT3}
\end{eqnarray}
$y = \log(\OIII\,\lambda5007/\Hbe)$, $x_1 = \log(\NII\,\lambda6583/\Hae)$, 
$x_2 = \log(\SII\,\lambda\lambda6716, 6731/\Hae)$, and $x_3 = \log(\OI\,\lambda6300/\Hae)$.
These star formation--AGN boundaries are determined by considering photoionization
by extremely young stars. These classifications show that the majority of our samples
consist of star-forming galaxies. Erring on the side of caution, we consider galaxies
that satisfy any of the three BPT criteria as potential AGNs.
The possible AGNs in the \OIIIa-detected sample are MK01, MK02, MMT07, and MMT11.
For the \OIIIa-non-detected sample, the possible AGNs are MK10, MMT40, MMT43, MMT62,
MMT66, MMT69, MMT76, MMT89, Keck051, Keck063, Keck085, and Keck089. None of our
galaxies with \OI\ measurements are LINERs.

One limitation of these diagnostics is that they are unavailable in optical spectra
for our higher redshift galaxies ($z\gtrsim0.4$). To supplement our \OI\ measurements,
we use a variety of emission-line flux ratios (\OII/\OIII\ and
\OII/\NeIII\,$\lambda$3869), to determine whether any of our higher redshift galaxies
could harbor a LINER. Upon comparing our emission-line fluxes to SDSS DR7 LINERs,
we find that MMT03 is arguably a LINER.
We also illustrate in Figure~\ref{fig:BPT_MEx} the ``Mass--Excitation'' (MEx) diagram
\citep{jun14}, which substitutes stellar mass (see Section~\SSED\ of \citetalias{MACTI})
for \NII\,$\lambda$6583/\Ha. This figure provides further support that the majority
of our samples consist of star-forming galaxies. Two galaxies (Keck038 and Keck099)
in the \OIIIa-non-detected sample might be AGNs. However, because the MEx diagnostic
is affected by evolution in the \MZ\ relation \citep[see Section~\ref{sec:MZ};][]{jun14},
we do not consider these sources as likely AGNs. We observe a turnover in the
MEx plot at $M_{\star}\sim10^8$ \Msun, which is due to the lower metal abundances
($\OHm\lesssim8.0$) in lower stellar mass galaxies.

To summarize, we suspect that 5 of the \Ndet\ \OIIIa-detected galaxies (8\%) and 12
of \Nrel\ \OIIIa-non-detected galaxies (12\%) are LINERs or AGNs. While these
AGN/LINER fractions are low, we note that other narrow-band studies, such as
\cite{rey15}, have also found low AGN/LINER contamination fractions (8\%).


\section{RESULTS}
\label{sec:Results}


\subsection{Extremely Metal-poor Galaxies}
We have identified a total of \NXMPG\ {\it extremely} metal-poor galaxies with \OH\
$\leq$ 7.69 (i.e., less than 10\% of solar). This is the largest extremely metal-poor
galaxy sample at $z\gtrsim0.1$.
Keck06 is our most metal-poor galaxy with $\OHm = 7.23^{+0.11}_{-0.14}$ (3\% of solar
metallicity). This is similar to I Zw 18, which is the most metal-deficient galaxy
known in the local universe.

We find that 24\% of our \OIIIa-detected galaxies are extremely metal-poor; this is
far higher than the 4\% of \OIIIa-detected galaxies in SDSS that are extremely
metal-poor \citep{izo06b}. This is presumably attributable to a combination of
redshift evolution (lower metallicity toward higher redshift; see
Section~\ref{sec:MZ}) and selection effects, because our sample is focused on
lower-mass galaxies ($\lesssim10^9$ \Msun) that tend to have lower metallicity.
If the extremely metal-poor galaxy fraction increases toward even lower masses, it
is possible that a substantial minority of local galaxies---{\it by number}---are
extremely metal-poor, even though their total mass is only a small fraction of the
current total stellar mass in the universe.
We suggest that future selections of extremely metal-poor galaxies should either
use narrow-band imaging or grism spectroscopy. This is more efficient observationally
than a brute-force approach within a magnitude-limited survey. For example, the
DEEP2 survey \citep{new13}, which targeted $R_{\rm AB}\lesssim24$ galaxies, has
identified only two extremely metal-poor galaxies at $z\sim0.8$ from a sample of
28 \OIIIa-detected galaxies \citep{ly15}.


\subsection{Specific Star Formation Rates and the \Mstar--SFR Relation}

\begin{figure*}
  \epsscale{1.1}
  \plotone{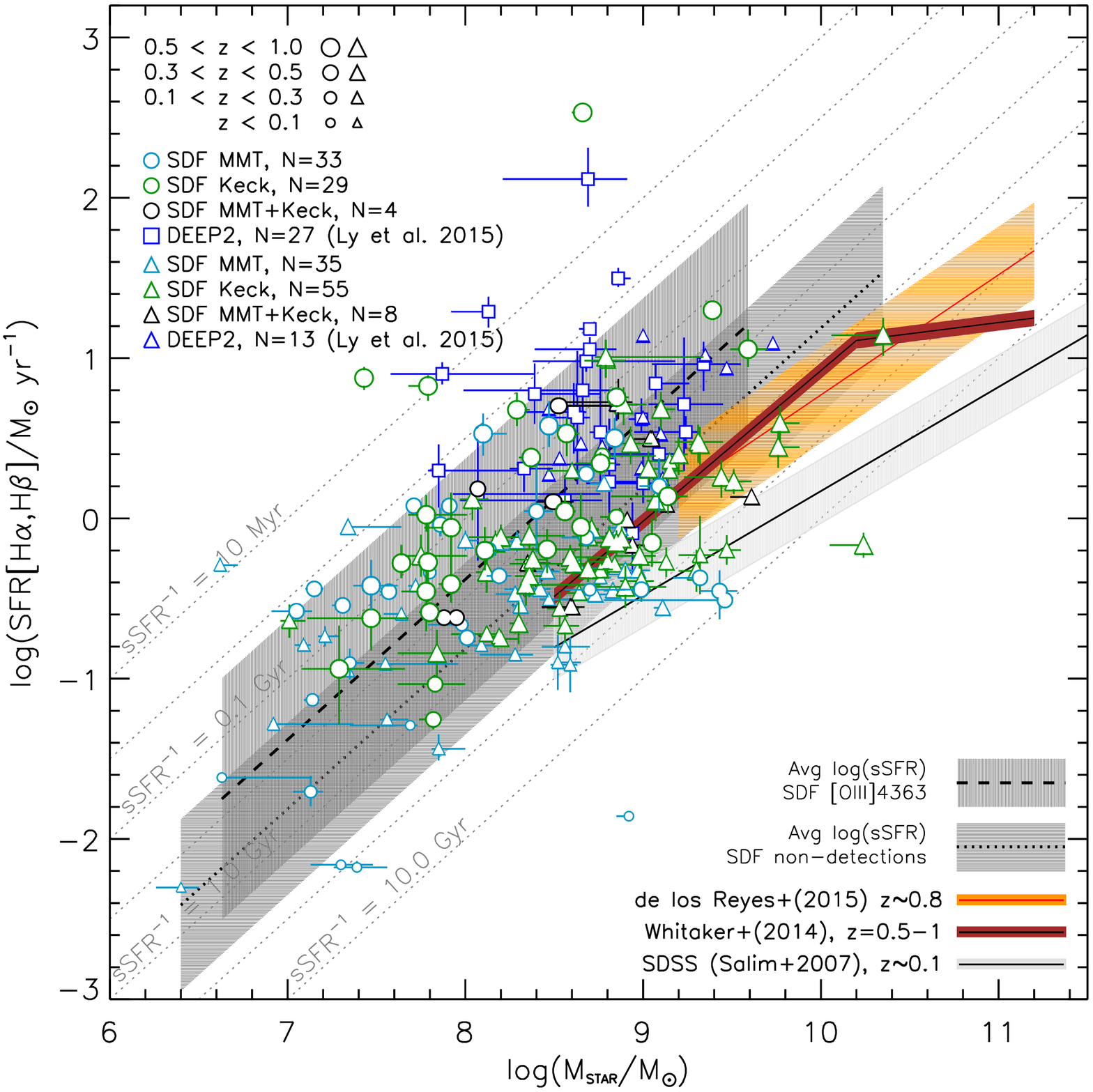}
  \caption{Dust-corrected SFR as a function of stellar mass for our SDF galaxies.
    The stellar masses are obtained from SED fitting (Section~\SSED\ of
    \citetalias{MACTI}). The SFRs are determined from either \Ha\ or \Hb\ luminosities
    (see Tables~\TSFR\ and \TSFRND\ of \citetalias{MACTI}), which are sensitive to
    a timescale of $\lesssim$10 Myr. The circles and triangles show galaxies with
    \OIIIa\ detections and \OIIIa\ non-detections, respectively. Light blue, green,
    and black points show our SDF galaxies observed with MMT, Keck, and both
    telescopes. The symbol size increases with redshift. In addition, we overlay the
    metal-poor DEEP2 galaxies from \cite{ly15} as dark blue squares and dark blue
    triangles. Gray dotted diagonal lines show different timescales of star formation,
    inverse specific SFR, or sSFR$^{-1}$. The averages of the inverse sSFRs for our
    \OIIIa-detected and \OIIIa-non-detected galaxies are \sSFRt\ and 650 Myr, shown
    by the dashed black line and dotted line, respectively. The \Mstar--SFR relations
    of \cite{sal07}, \cite{whi14}, and \cite{rey15} at $z=0.1$, $z=0.5$--1, and
    $z=0.8$, are illustrated by the gray, brown, and orange bands, respectively,
    with the dispersion in sSFR illustrated by the shaded regions. Our
    \OIIIa-non-detected galaxies are consistent with the \Mstar--SFR relations at
    similar redshift, whereas our \OIIIa-detected galaxies tend to lie about a factor
    of $\approx$3 above the \Mstar--SFR relation. A broad dispersion in sSFR suggests
    that \OIIIa\ can be detected in ``typical'' star-forming galaxies at $z\lesssim1$.}
  \label{fig:SFR_Mass}
\end{figure*}

In Figure~\ref{fig:SFR_Mass}, we compare our dust-corrected instantaneous SFRs from
\Ha\ or \Hb\ luminosities against stellar masses determined from spectral energy
distribution (SED) fitting, to locate our galaxies on the \Mstar--SFR relation and
to compare against other star-forming galaxies at $z\lesssim1$. While the SFRs for
the \OIIIa-detected and \OIIIa-non-detected galaxies are modest ($\approx$0.1--10
\Msun\ yr$^{-1}$), their stellar masses are 1--2 dex lower than galaxies generally
observed at $z\sim1$. Therefore, we find that our emission-line galaxies are all
undergoing relatively strong star formation.
The specific SFRs (SFR per unit stellar mass, SFR/\Mstar; hereafter sSFR) that we
measure are between $10^{-10.8}$ yr$^{-1}$ and $10^{-6.1}$ yr$^{-1}$ with an average
of \sSFRA\ yr$^{-1}$ for the \OIIIa-detected sample, and between
$10^{-10.4}$ yr$^{-1}$ and $10^{-6.9}$ yr$^{-1}$ with an average of $10^{-8.8}$ yr$^{-1}$
for the \OIIIa-non-detected sample. These averages are illustrated in
Figure~\ref{fig:SFR_Mass} by the dashed black line and dotted black line for the
\OIIIa-detected and \OIIIa-non-detected samples, respectively. The gray shaded
regions indicate the 1$\sigma$ dispersion in sSFR for the samples.

These sSFRs are enhanced by 0.25--4.0 dex above the \Mstar--SFR relation for $z\sim0$
SDSS galaxies \citep{sal07}. Extrapolating the \Mstar--SFR relation of \cite{whi14}
and \cite{rey15} toward lower stellar mass, we find that the sSFRs of our
emission-line galaxies are $\approx$0.0--3.0 dex higher than ``typical'' galaxies
at $z\sim0.45$--0.85.
While our sample is biased toward stronger star formation activity (see
Section~\ref{sec:bias}), 44\% of \OIIIa-detected and \OIIIa-non-detected galaxies
lie within $\pm$0.3 dex (i.e., 1$\sigma$) of the $z\sim0.8$ \Mstar--SFR relation of
\cite{rey15} and an additional 17\% of our samples lie below the \Mstar--SFR relation
by more than 0.3 dex (see Figure~\ref{fig:SFR_Mass}).
For comparison, our previous \OIIIa-detected study \citep{ly14}, which had shallower
spectroscopy by a factor of $\sim$2, yielded a significant sSFR offset of $\approx$1
dex on the \Mstar--SFR relation from typical star-forming galaxies at $z\leq1$.
The deeper observations of \Sname\ result in a lower sSFR by $\approx$0.5 dex. We
also illustrate the \cite{ly15} \OIIIa-selected metal-poor sample from DEEP2 as
blue squares and triangles in Figure~\ref{fig:SFR_Mass}. This DEEP2 sample consists
of galaxies with higher SFR activity than our \OIIIa-detected and \OIIIa-non-detected
samples, which is in part due to the shorter integration time of DEEP2 (1 hr) than
\Sname\ (2 hr). It can also be seen that the \Sname\ sample extends to lower stellar
mass by $\approx1$ dex at $z\sim1$ than \cite{ly15}.
%


\subsection{Lower Redshift Analogs to $z\gtrsim2$ Galaxies}
\begin{figure}
  \epsscale{1.1}
  \plotone{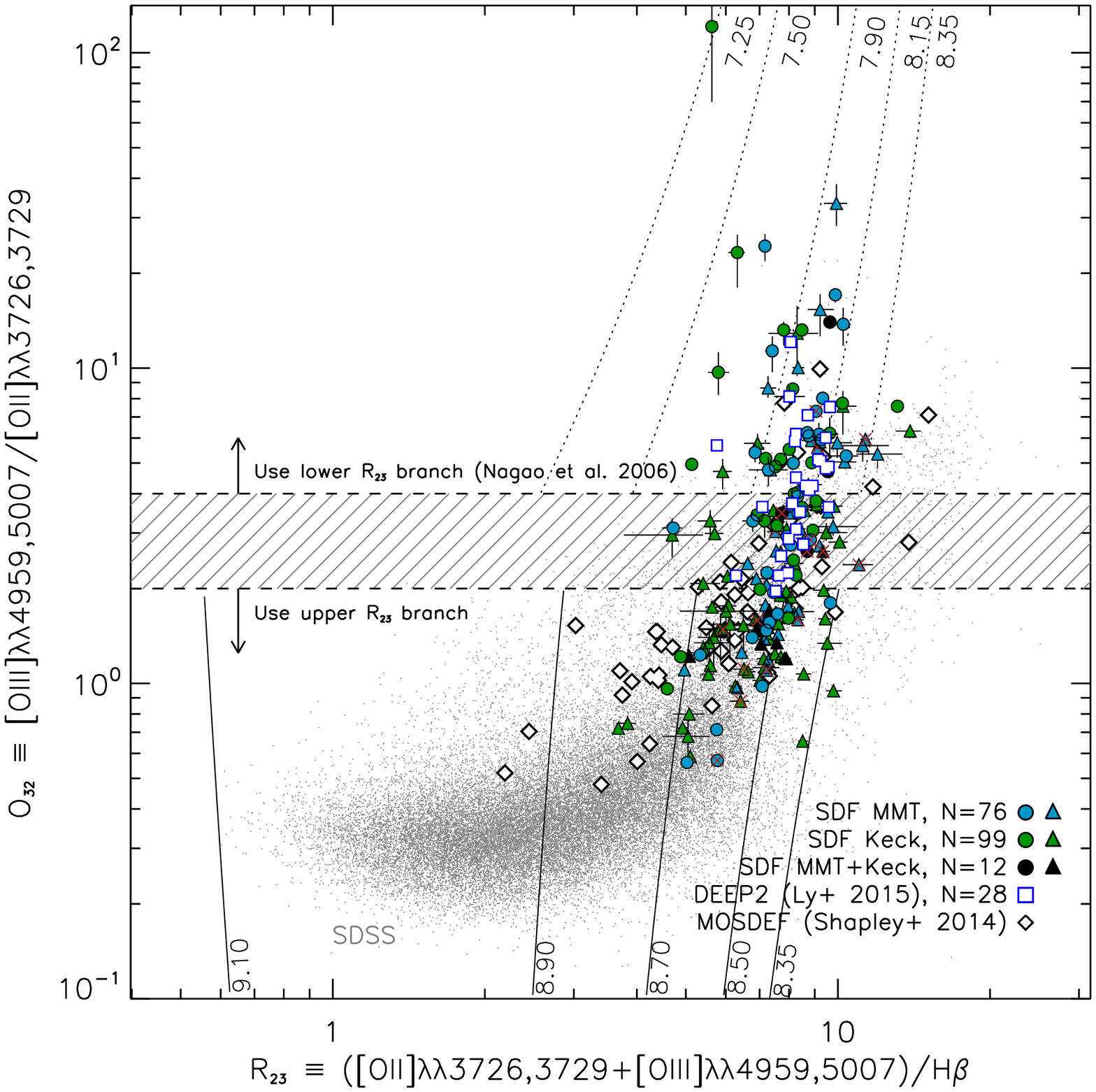}
  \caption{Metallicity-sensitive ($R_{23}$) and ionization parameter-sensitive ($O_{32}$)
    emission-line ratios for SDF \OIIIa-detected (circles) and \OIIIa-non-detected
    (triangles) samples from MMT (light blue), Keck (green), and both (black). SDF
    galaxies with brown crosses indicate possible AGNs and LINERs (see
    Section~\ref{sec:AGN}). DEEP2 \OIIIa-detected galaxies are also overlaid as dark
    blue squares, and local galaxies from SDSS are shown by the gray points.
    Metal-poor galaxies from both SDF and DEEP2 lie along a ``ridge'' consisting
    of high-$R_{23}$ and high-$O_{32}$ values. Typical $z\sim2$ galaxies
    \citep[black diamonds;][]{sha15} are found along this same ``ridge,'' suggesting
    that $z\approx0.2$--1 metal-poor galaxies are analogous to $z\gtrsim2$ star-forming
    galaxies. For illustration purposes, photoionization model tracks from \cite{m91}
    are overplotted for metallicities between \OH\ = 7.25 and \OH\ = 9.1. Solid
    (dotted) curves are for metallicities on the upper (lower) $R_{23}$ branch. Based
    on the empirical relations of \cite{nag06}, the dashed horizontal lines distinguish
    between the upper and lower $R_{23}$ branches with a region of ambiguity
    (gray line-filled region).}
  \label{fig:R23_O32}
\end{figure}

We illustrate in Figure~\ref{fig:R23_O32} the $R_{23}$ and $O_{32}$ strong-line
ratios \citep{pag79}:
\begin{eqnarray}
  R_{23} &\equiv& \frac{\OII\,\lambda\lambda3726,3729 + \OIII\,\lambda\lambda4959,5007}{\Hbe},{\rm~and}\\
  O_{32} &\equiv& \frac{\OIII\,\lambda\lambda4959,5007}{\OII\,\lambda\lambda3726,3729}.
\end{eqnarray}
We compare our \OIIIa-detected and \OIIIa-non-detected samples to typical $z\sim2$
star-forming galaxies identified by the ``MOSDEF'' survey \citep[black diamonds in
  this figure;][]{kri15,sha15}. We find that our metal-poor galaxies have similar
interstellar properties (low metallicity, high ionization parameter) to the higher
redshift galaxy population, suggesting that we have identified low-$z$ analogs to
$z\gtrsim2$ galaxies.
Specifically, the MOSDEF survey detects \OIII$\lambda$5007 at S/N = 100 of
$\approx3\times10^{-16}$ erg s$^{-1}$ cm$^{-2}$ or a line luminosity of
$4\times10^{42}$ erg s$^{-1}$ at $z=1.5$, $1.3\times10^{43}$ erg s$^{-1}$ at $z=2.35$,
and $3\times10^{43}$ erg s$^{-1}$ at $z=3.35$ \citep{kri15}. As illustrated in
Figure~\FLum\ of \citetalias{MACTI}, the average \OIII\ luminosity of the
\OIIIa-detected sample from \Sname\ is 1.3--2.1 dex lower than the sensitivity of
MOSDEF. Because the MOSDEF survey integrated for $\sim$1--2 hr, the \OIIIa\ emission
for galaxies at $z\gtrsim1.3$ would require at least $\sim$100 hours of Keck/MOSFIRE
observations for individual S/N = 3 detections.


\subsection{The Mass--Metallicity Relation}
\label{sec:MZ}
\begin{figure*}
  \epsscale{1.1}
  \plotone{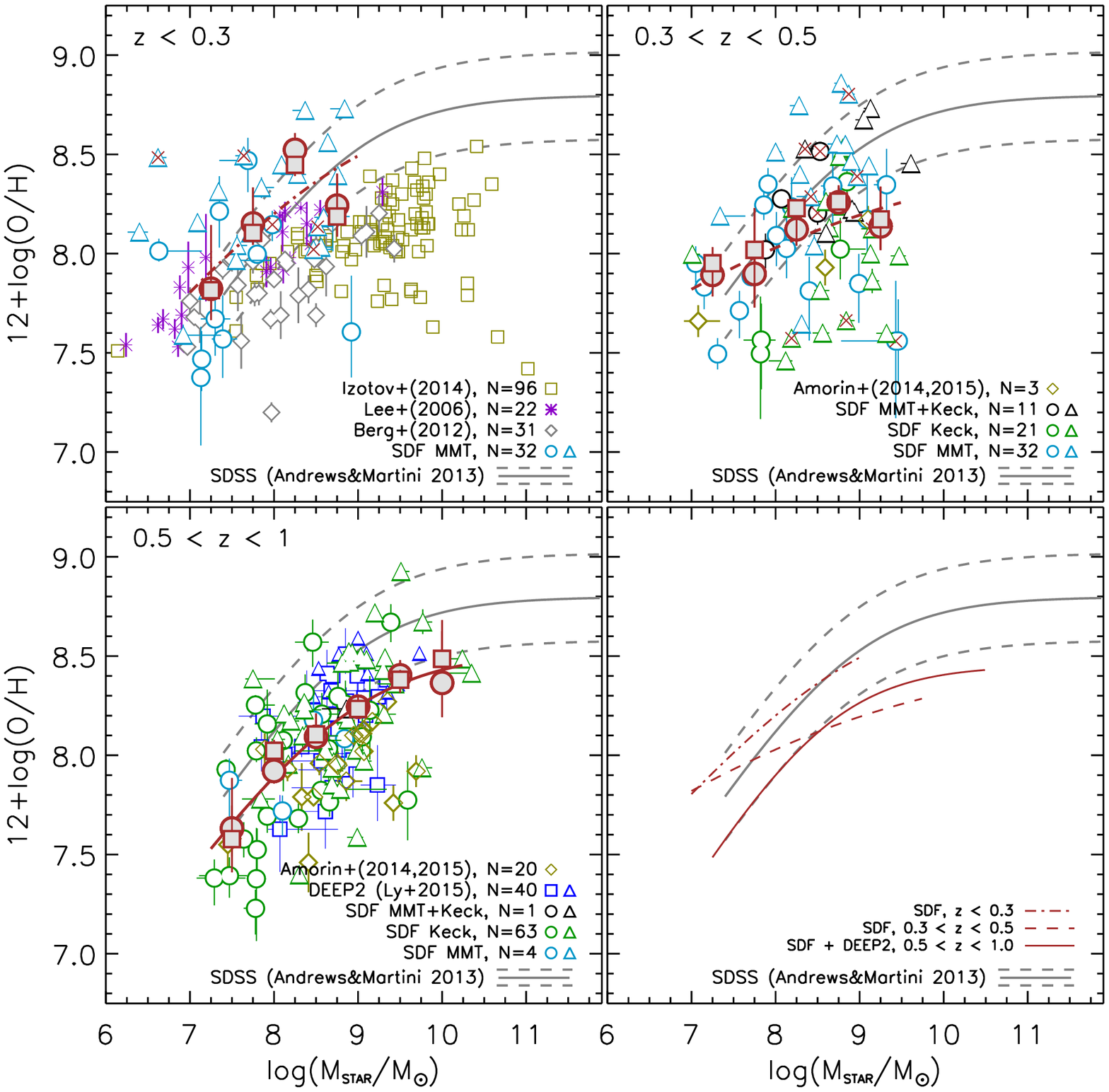}
  \caption{O/H abundance as a function of stellar mass for three redshift bins:
    $z\leq0.3$ (upper left), $0.3< z\leq0.5$ (upper right), and $0.5< z\leq1$
    (lower right). The light blue, green, and black symbols are SDF galaxies from
    MMT, Keck, and both, respectively. Circles (triangles) illustrate
    \OIIIa-detected (\OIIIa-non-detected) galaxies. DEEP2 galaxies from \cite{ly15}
    with \OIIIa\ detections (\OIIIa\ non-detections) are overlaid in the lower left
    panel as dark blue squares (triangles). Additional samples from
    \cite[purple asterisks]{lee06}, \cite[dark gray diamonds]{berg12},
    \cite[olive diamonds]{amo14,amo15}, and \cite[olive squares]{izo14} are shown
    for comparison. Large brown symbols show averages (circles) and median (squares)
    in bins of stellar mass computed from the SDF and DEEP2 samples. The averages are
    fitted with the three-parameter curve (Equation~(\ref{eqn:MZ})), which is shown
    by the dark brown curves. We compare our \MZ\ relation to SDSS galaxies from
    \citetalias{and13}, which is shown by a solid gray line, with gray dashed lines
    enclosing the $\pm$1$\sigma$. Here, the scatter of \citetalias{and13} is not
    the true intrinsic scatter from individual galaxies. Rather, it reflects the
    dispersion for stacked spectra in various \Mstar--SFR bins. This $\sigma$ is
    likely to be larger than the intrinsic scatter because it is weighted more toward
    high-SFR outliers (fewer galaxies are available in these bins; B. Andrews 2013,
    private communication). Brown crosses indicate SDF galaxies that are possible
    AGNs and LINERs (see Section~\ref{sec:AGN}), which are excluded from average and
    median measurements. For comparison purposes, the lower right panel illustrates
    the best fit for each redshift bin.}
  \label{fig:MZ}
\end{figure*}

\begin{figure}
  \epsscale{1.1}
  \plotone{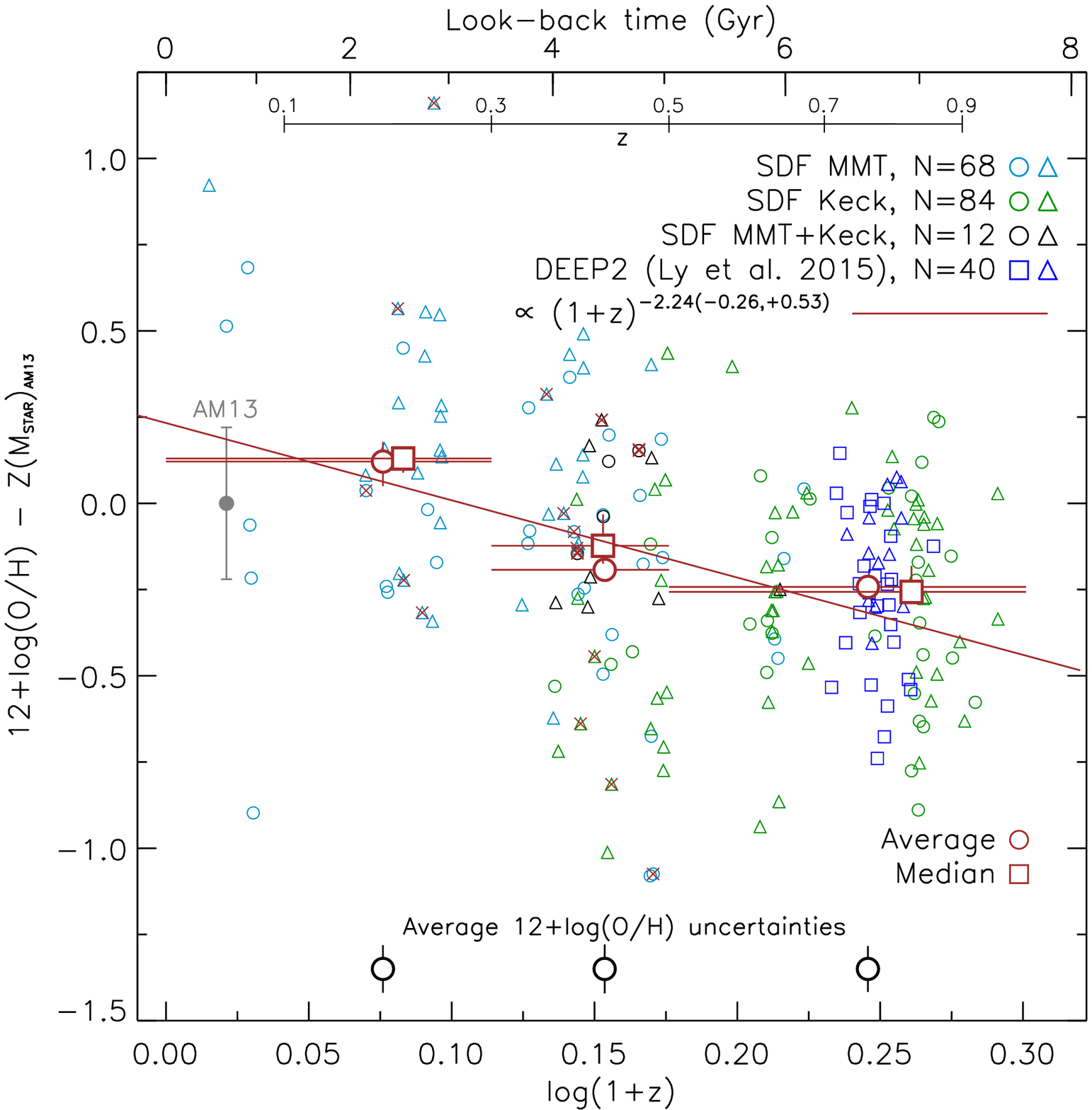}
  \caption{O/H abundance as function of redshift or look-back time. Here, the
    abundances are illustrated relative to the local \MZ\ relation of
    \citetalias{and13}. The light blue, green, and black symbols are SDF galaxies
    from MMT, Keck, and both, respectively. Circles (triangles) illustrate galaxies
    with \OIIIa\ detections (upper limits). DEEP2 galaxies are overlaid in dark blue
    with \OIIIa\ detections (squares) and upper limits (triangles). Large brown
    symbols show the average (circles) and median (squares) computed from the SDF
    and DEEP2 samples. The uncertainties on the median and average values are
    determined by statistically bootstrapping: random sampling with replacement,
    repeated 10,000 times. The best fit to the average measurements is shown by
    the brown line, which shows a strong redshift dependence, $(1+z)^{\zslope}$.
    To demonstrate that the majority of the observed scatter is not due to
    measurement uncertainties, we illustrate the average \OH\ uncertainties in each
    redshift bin (for the SDF sample) using the black circles near the bottom of the
    figure. Brown crosses indicate SDF galaxies that are possible AGNs and LINERs
    (see Section~\ref{sec:AGN}), which are excluded from average and median
    measurements.}
  \label{fig:MZ_offset}
\end{figure}

\begin{figure}
  \epsscale{1.05}
  \plotone{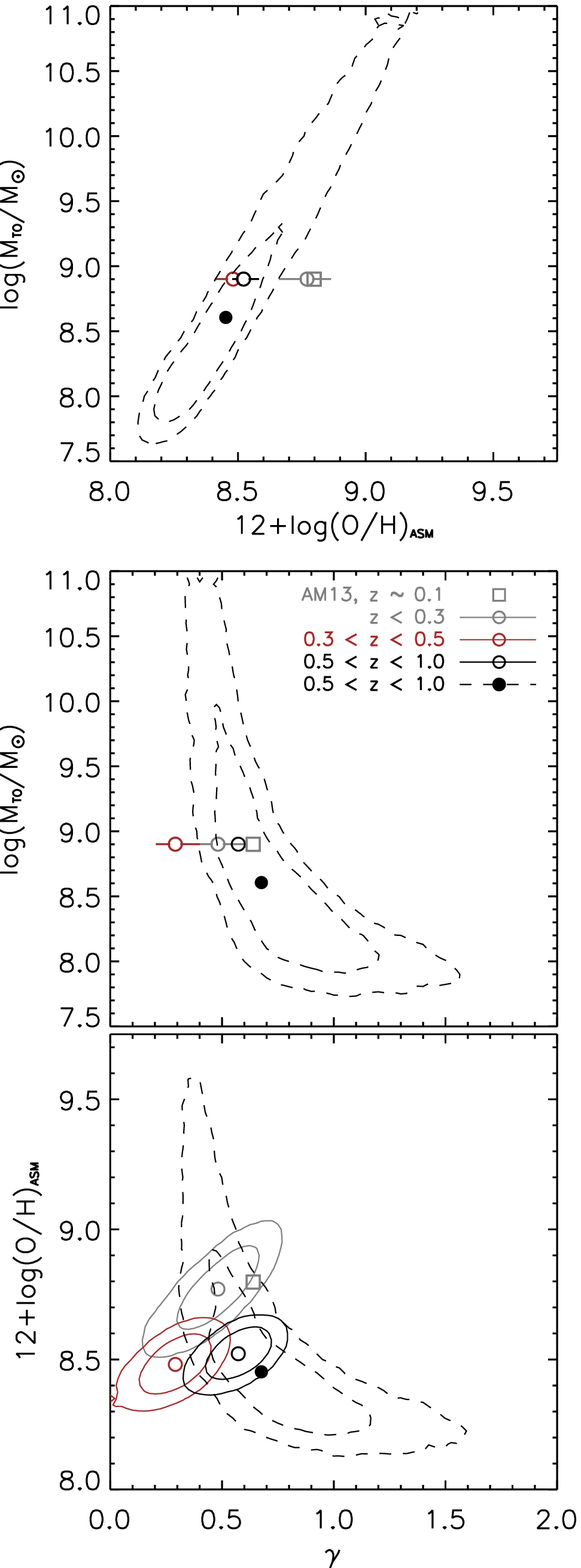}
  \caption{Confidence contours (or error bars) for the best fit to the \MZ\ relations
    with a three-parameter function of a turnover mass ($M_{\rm TO}$), an asymptotic
    metallicity at high mass ($\OHm_{\rm asm}$), and the low-mass end slope ($\gamma$)
    (see Equation~(\ref{eqn:MZ})). Error bars illustrate 68\% confidence, and
    confidence contour levels for 68\% and 95\% are also shown. Our fitting results
    for the \MZ\ relation are shown by the circles for $z<0.3$ (gray), $0.3 < z < 0.5$
    (brown), and $0.5 < z < 1$ (black). Unfilled circles with solid lines are those
    with $\log(M_{\rm TO}/M_{\sun})$ fixed to 8.901, the local value of \citetalias{and13}.
    For our highest redshift bin, the filled circles with dashed contours illustrate
    the fitting results with a free $\log(M_{\rm TO}/M_{\sun})$. For comparison, we
    overlay the results of \citetalias{and13} by the gray squares.}
  \label{fig:MZ_fit_contours}
\end{figure}

We illustrate in Figure~\ref{fig:MZ} the dependence of oxygen abundance on stellar
mass in three redshift bins, $z\leq0.3$, $z=0.3$--0.5, and $z=0.5$--1. In
Figure~\ref{fig:MZ_offset}, we compare the \OIIIa-detected and
\OIIIa-non-detected samples from this paper and the DEEP2 \OIIIa-detected and
\OIIIa-non-detected samples from \cite{ly15} against the \citet[hereafter AM13]{and13}
\MZ\ relation of the form:
\begin{equation}
  12+\log{\left({\rm O/H}\right)} = 12+\log{\left({\rm O/H}\right)}_{\rm asm} -
  \log{\left[1 + \left(\frac{M_{\rm TO}}{M_{\star}}\right)^{\gamma}\right]},
  \label{eqn:MZ}
\end{equation}
where \OH$_{\rm asm}$ is the asymptotic metallicity at the high mass end, $M_{\rm TO}$
is the turnover mass or ``knee'' in the \MZ\ relation, and $\gamma$ is the slope
of the low-mass end. This formalism is consistent with \cite{mou11} in describing
the \MZ\ relation, and provides an intuitive understanding for the shape of the
\MZ\ relation. For $z\sim0.1$, \citetalias{and13} find a best fit of
\OH$_{\rm asm}=8.798$, $\log(M_{\rm TO}/M_{\sun})=8.901$, and $\gamma=0.640$.
At a given stellar mass, these emission-line selected samples are (on average)
offset in \OH\ by 0.13$^{+0.06}_{-0.07}$ dex at $z\leq0.3$, --0.17$^{+0.07}_{-0.03}$
dex at $z=0.3$--0.5, and --0.24$\pm$0.03 dex at $z=0.5$--1.
This demonstrates a moderate evolution in the \MZ\ relation of:
\begin{equation}
  \OHm - Z(M_{\star})_{{\rm AM13}} = A + B \log(1+z),
\end{equation}
where $A = 0.29^{+0.04}_{-0.13}$ and $B = \zslope$. To better understand this
evolution, we compute the average and median in each stellar mass bin, provided in
Table~\ref{tab:binned_mz}, and shown as brown squares (average) and circles
(median) in Figure~\ref{fig:MZ}.
We then fit the averages with Equation~(\ref{eqn:MZ}) using MPFIT \citep{mar09}.
The fitting is repeated 10,000 times with each fit using the bootstrap approach to
compute the average in each stellar mass bin. The best-fitting results are provided
in Table~\ref{tab:mz_best_fit} and the confidence contours are illustrated in
Figure~\ref{fig:MZ_fit_contours}.

With only four or five stellar mass bins below \Mstar\ $\sim10^9$ \Msun\ for our
two lowest redshift bins ($z<0.5$), fitting results are poorly constrained with
all three parameters free. Specifically, the turnover mass ($M_{\rm TO}$) and the
asymptotic metallicity (\OH$_{\rm asm}$) require measurements at higher stellar
masses. In addition, with smaller sample sizes for these redshifts the best fits
can easily be affected by a small number of outliers (e.g., the highest mass bin
for $0.3<z<0.5$ has a large number of metal-poor galaxies). For these reasons, we
fixed $M_{\rm TO}$ to the local value obtained by \citetalias{and13}:
$10^{8.901}$ \Msun. We find that the best-fit result to the SDF \MZ\ relation at
$z<0.3$ is consistent with \citetalias{and13} within measurement uncertainties.
At $0.3<z<0.5$, the best fit yields a lower \OH$_{\rm asm}$\ by $\approx0.3$ dex.
At $z=0.5$--1, our observations extend to \Mstar\ $\sim10^{10}$ \Msun\ and there
are significantly more galaxies to better constrain the shape. Thus, we allow
$M_{\rm TO}$ to be a free parameter in the fitting, in addition to adopting the
local value. Our best fits to the \MZ\ relation at $z=0.5$--1 indicate that the
shape of the \MZ\ relation remains unchanged at $z\sim1$, but with lower a
\OH$_{\rm asm}$ by $\approx$0.30 dex (i.e., a lower metallicity at all stellar masses).


\input tab1.final.tex

\input tab2.final.tex


\subsection{Dependence on SFR}
Several observational and theoretical investigations have proposed that the \MZ\
relation has a secondary dependence on the SFR
\citep[see e.g.,][and references therein]{ell08,lar10,man10,dave11,lil13,sal14}.
Specifically, the lower abundances at higher redshift may be explained by higher sSFR,
such that there is a non-evolving (i.e., ``fundamental'') relation \citep{lar10,man10}.
To test this relation, we adopt a non-parametric method of projecting the \MZ--SFR
relation in various two-dimensional spaces.

\begin{figure*}
  \epsscale{1.1}
  \plotone{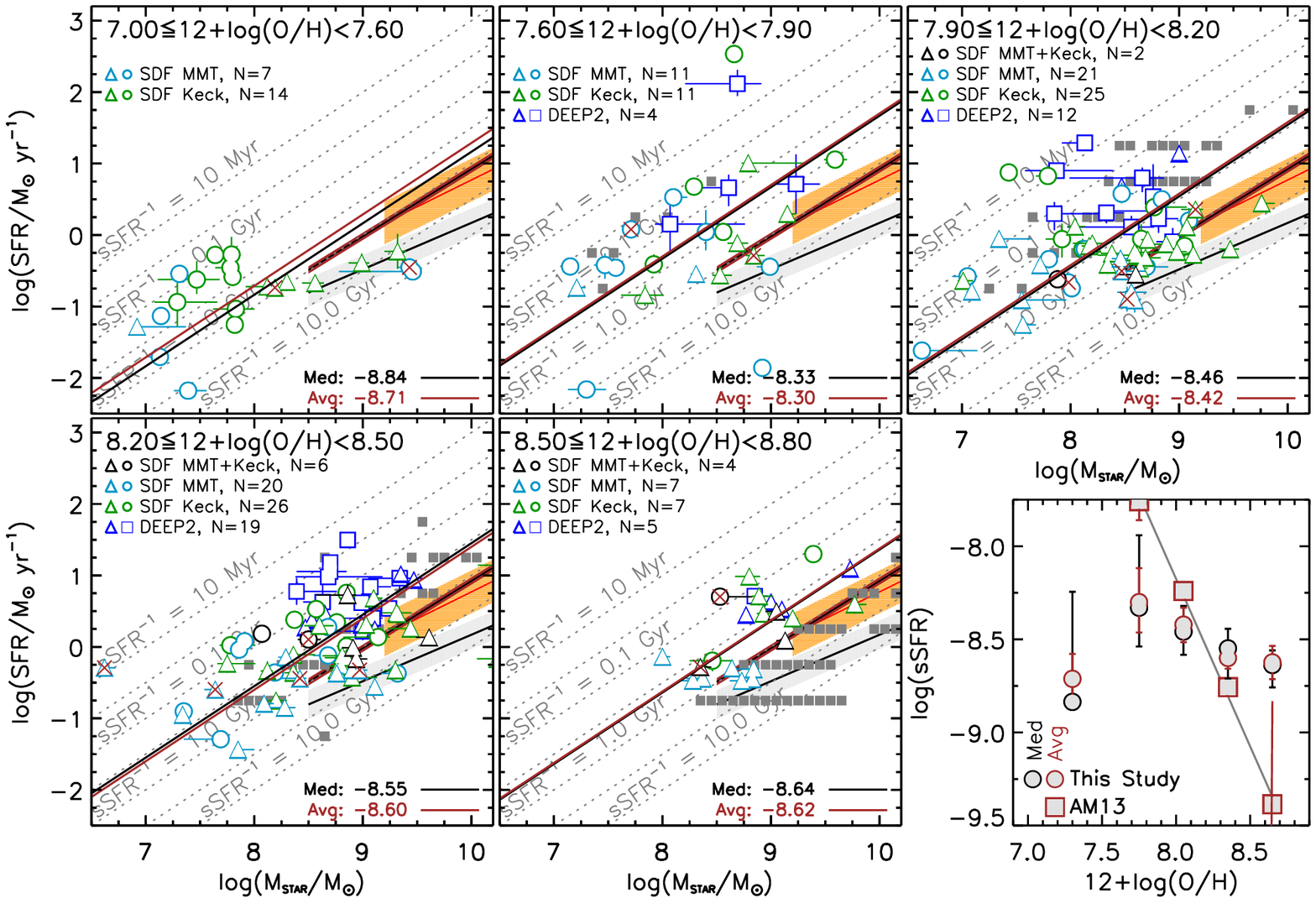}
  \caption{The correlation between dust-corrected SFR and stellar mass, for a given
    metallicity. While this figure is similar to Figure~\ref{fig:SFR_Mass}, each
    panel shows galaxies in different metallicity bin, ranging from the lowest O/H
    in the upper left to the highest in the lower middle panel. The stellar masses
    are obtained from SED fitting (Section~\SSED\ of \citetalias{MACTI}). The SFRs
    are determined from either the \Ha\ or \Hb\ luminosity (see Tables~\TSFR\ and
    \TSFRND\ of \citetalias{MACTI}), which are sensitive to a timescale of
    $\lesssim$10 Myr. Galaxies with detections of \OIIIa\ are shown as circles,
    while triangles show the \OIIIa-non-detected samples. The SDF galaxies observed
    by Keck are shown in green, while light blue points are those observed by MMT.
    Those observed by both telescopes are plotted as black circles or triangles.
    The DEEP2 galaxies are overlaid as dark blue squares and triangles. The average
    and median sSFR to the SDF and DEEP2 datasets are given by the brown and black
    lines, respectively, and are shown in the lower right panel against \OH\ as
    black (median) or brown (average) circles. The uncertainties on the median and
    average values are determined by statistically bootstrapping: random sampling
    with replacement, repeated 10,000 times. For comparisons, we also overlay the
    \citetalias{and13} stacked samples as gray squares in each metallicity bin with
    average sSFR as brown squares in the lower right panel. The \Mstar--SFR relation
    determined by \cite{sal07}, \cite{rey15}, and \cite{whi14} are plotted as gray
    ($z\sim0$), orange ($z\sim0.8$), and brown ($z=0.5$--1) bands, respectively.
    Lines of constant inverse specific SFR (sSFR$^{-1}$) are shown by the dotted
    gray lines, with corresponding timescale. As illustrated in the lower right
    panel, the sSFR increases toward lower metallicity, but at rate that is shallower
    than the local results of \citetalias{and13} (gray solid line). Brown crosses
    indicate SDF galaxies that are possible AGNs and LINERs (see
    Section~\ref{sec:AGN}), which are excluded from average and median measurements.}
  \label{fig:FMR_MS}
\end{figure*}

First, we illustrate in Figure~\ref{fig:FMR_MS} the location on the \Mstar--SFR plane
\citep{noe07,sal07} for galaxies in five different metallicity bins. We then compute
the average and median sSFR for each bin. These are shown as brown and black solid
lines, respectively, in each panel, and are summarized in the lower right panel.
The hypothesis we are testing is whether, for a given stellar mass, galaxies shift
toward higher SFRs as metallicity decreases. Our results show that indeed the sSFR
is lower for higher values of $\log({\rm O/H})$, except at the lowest abundance bin.
For our lowest abundance bin, Figure~\ref{fig:FMR_MS} shows that the distribution
in sSFR is skewed (as evident by a $\sim$0.2 dex difference between the median and
average values) by a small number of higher mass galaxies with low sSFR.
The $\log({\rm sSFR})$--$\log{({\rm O/H})}$ slope that we measure is shallower
than \citetalias{and13}, --0.30 vs. --1.80.
Specifically, the greatest difference in sSFR of $\sim$0.8 dex is at high metallicities.
This difference is likely caused by a bias in our survey toward higher sSFR because
metal-rich galaxies with low SFRs will not have \OIIIa\ detections and will fall
below our flux limit cuts adopted for the \OIIIa-non-detected sample.
We defer a discussion on selection bias to Section~\ref{sec:bias}.


Next, we consider a projection first adopted by \cite{sal14}: O/H as a function of
the vertical offset on the \Mstar--SFR relation. The offset, defined as \DsSFR,
measures the excess of star formation relative to ``normal'' galaxies of the same
stellar mass and redshift.
To facilitate comparisons with the local results of \citetalias{and13}, we use the
\cite{sal07} $z\sim0.1$ \Mstar--SFR relation as our reference relation:
\begin{equation}
  \log{\left({\rm SFR}/M_{\sun}~{\rm yr}^{-1}\right)} = 0.65\log{\left(M_{\star}/M_{\sun}\right)} - 6.33.
\end{equation}

\begin{figure*}
  \epsscale{1.1}
  \plotone{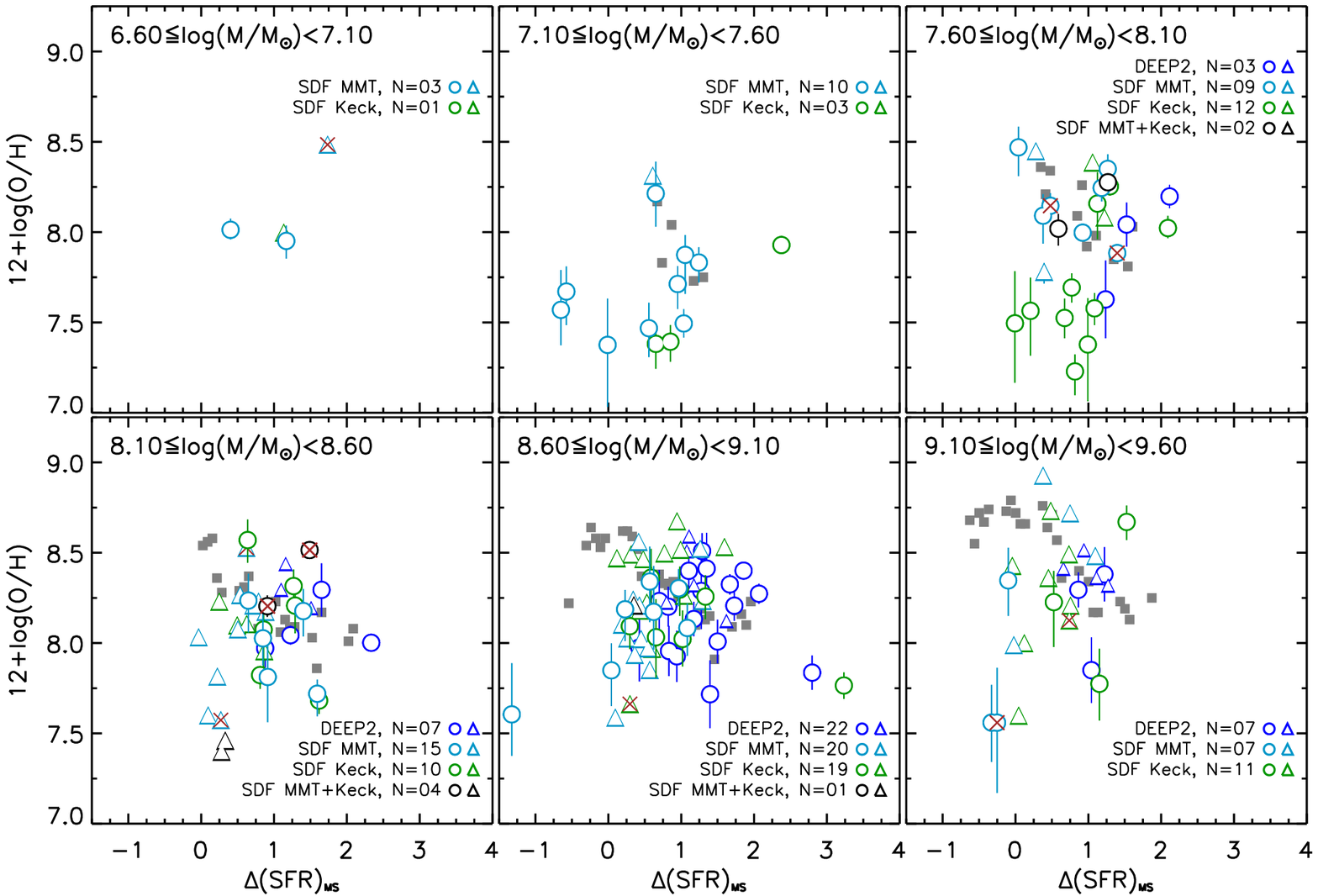}
  \caption{Oxygen abundance as a function of SFR offset from the local star-forming
    \Mstar--SFR relation, \DsSFR\ \citep{sal07}. Individual panels show galaxies in
    six different bins of stellar mass, increasing from the smallest dwarfs (upper
    left) to a tenth of typical massive galaxies (lower right). The circles
    show \OIIIa-detected galaxies, while the triangles show \OIIIa-non-detected
    galaxies. The light blue points represent MMT observations of SDF galaxies, while
    the green points represent Keck data. The black points show SDF galaxies observed
    with both telescopes, and the dark blue squares and triangles are DEEP2 galaxies.
    Brown crosses indicate SDF galaxies that are possible AGNs and LINERs (see
    Section~\ref{sec:AGN}). For comparison we overlay as filled gray squares the
    measurements of local galaxies from \citetalias{and13}, who found an inverse
    correlation in this diagram, where those galaxies with exceptionally strong
    sSFR's have lower metallicities. Our average values of abundances and \DsSFR\
    overlap with \citetalias{and13}. Our emission-line galaxies show a large scatter
    in this diagram, which is too large to see any significant inverse correlation
    between sSFR and metallicity, as \citetalias{and13} found.}
  \label{fig:FMR_Salim}
\end{figure*}

\noindent This metallicity--\DsSFR\ comparison is performed in different stellar mass
bins, and is illustrated in Figure~\ref{fig:FMR_Salim}. Here, we compare our sample to
\citetalias{and13}, which is indicated by filled gray squares. This figure illustrates
that our emission-line galaxy samples are qualitatively consistent with
\citetalias{and13}; however, the sSFR dependence is weak. Specifically, there is a
shallow inverse dependence at intermediate stellar masses
($8.1\leq\log(M_{\star}/M_{\sun})<8.6$), but no significant dependence in the remaining
stellar mass bins.
For these other stellar mass bins, there may be evidence for a \textit{positive}
metallicity--\DsSFR\ dependence; however, this is weak with significant dispersion
of $\approx$0.3 dex that is larger than measurement uncertainties.


\begin{figure*}
  \epsscale{1.1}
  \plotone{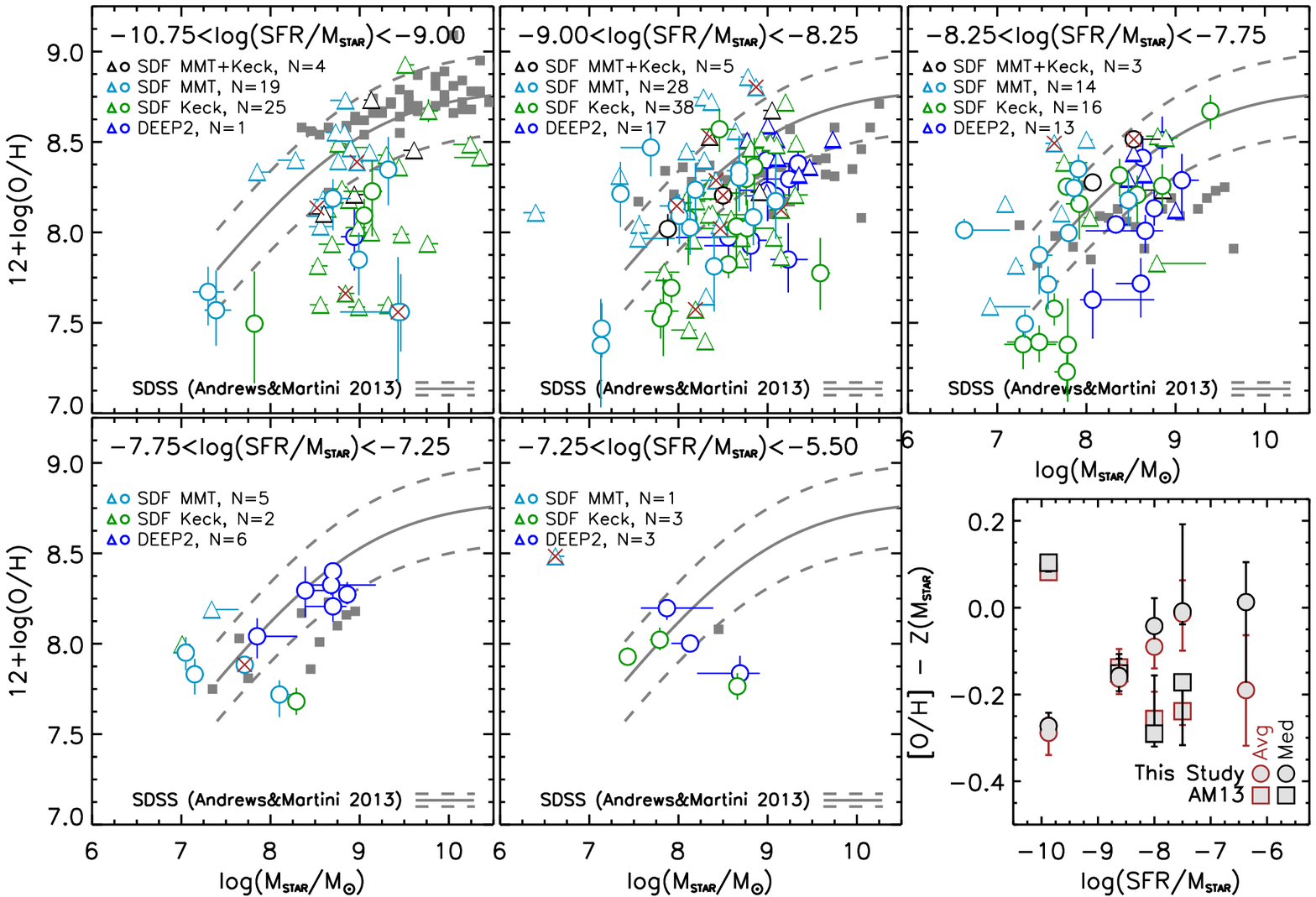}
  \caption{Oxygen abundance as a function of stellar mass in five different bins of
    $\log({\rm sSFR})$, increasing from low sSFRs (upper left) to the highest sSFRs
    (lower middle). The circles show SDF \OIIIa-detected galaxies, while the
    triangles show SDF \OIIIa-non-detected galaxies. The MMT, Keck, and the MMT+Keck
    samples from the SDF are shown in light blue, green, and black, respectively.
    The dark blue squares and triangles are DEEP2 galaxies from \cite{ly15}. For
    comparison we overlay as filled gray squares the measurements of local galaxies
    from \citetalias{and13}. The lower right panel illustrates the offset in
    metallicity against the \MZ\ relation of \citetalias{and13},
    ${\rm [O/H]} - Z(M_{\star})$, as a function of $\log({\rm sSFR})$. Average and
    median values are shown in brown and black, respectively, with our measurements
    in circles and local measurements in squares. The uncertainties on the median
    and average values are determined by statistically bootstrapping: random
    sampling with replacement, repeated 10,000 times.}
  \label{fig:FMR_MZ}
\end{figure*}

The last projection that we consider is how the \MZ\ relation depends on sSFR. This
is illustrated in Figure~\ref{fig:FMR_MZ} in five different sSFR bins from
$\log({\rm sSFR}/{\rm yr}^{-1}) = -9.8$ to --6.4. Similar to Figures~\ref{fig:FMR_MS}
and \ref{fig:FMR_Salim}, we overlay the \citetalias{and13} sample as filled gray
squares. The lower right panel of Figure~\ref{fig:FMR_MZ} illustrates the median
(black points) and average (red points) metallicities relative to the
\citetalias{and13} \MZ\ relation (see Equation~(\ref{eqn:MZ})).
While we find good agreement with \citetalias{and13} at $-9.00<{\rm sSFR}<-8.25$, our
results are broadly inconsistent with theirs. Specifically, we find that the relative
offset on the \MZ\ relation increases with increasing sSFR. However, as discussed
earlier, our selection function misses metal-rich galaxies. This effect has the
largest impact for the lowest sSFR bin; the upper left panel of Figure~\ref{fig:FMR_MZ}
shows that metal-rich galaxies at \Mstar\ $\gtrsim10^9$ \Msun\ are not included in
our sample. The effect of this selection would shift the average and median
metallicities lower, possibly producing a false positive dependence. Because of the
inability to measure metallicity in metal-rich galaxies with low star formation
activity, the use of spectral stacking \citepalias[such as][]{and13} is necessary
to obtain average \Te-based abundances.
  

\section{DISCUSSION}
\label{sec:Disc}


\subsection{Selection Function of the Survey}
\label{sec:bias}

One concern with our spectroscopic survey is the selection bias of requiring the
detection of \OIIIa. Specifically, detection of this line primarily depends on the
electron temperature (or gas metallicity), which corresponds to
\begin{eqnarray}
  T_e(\OIII)  & = & a \left(-\log(\mathcal{R})-b\right)^{-c},{\rm~where}\\
  \mathcal{R} & \equiv & \frac{F(\OIII\,\lambda4363)}{F(\OIII\,\lambda\lambda4959,5007)},
\end{eqnarray}
$a$ = 13205, $b$ = 0.92506, and $c$ = 0.98062 \citep{nic14}, and the dust-corrected
SFR and redshift, which determine the emission-line fluxes.
At high SFRs, the probability of detecting \OIIIa\ is greater for a wide range of
metallicity. This range in metallicity reduces such that only metal-poor galaxies
with low SFRs can be detected in an emission-line flux limited survey.

\begin{figure*}
  \plotone{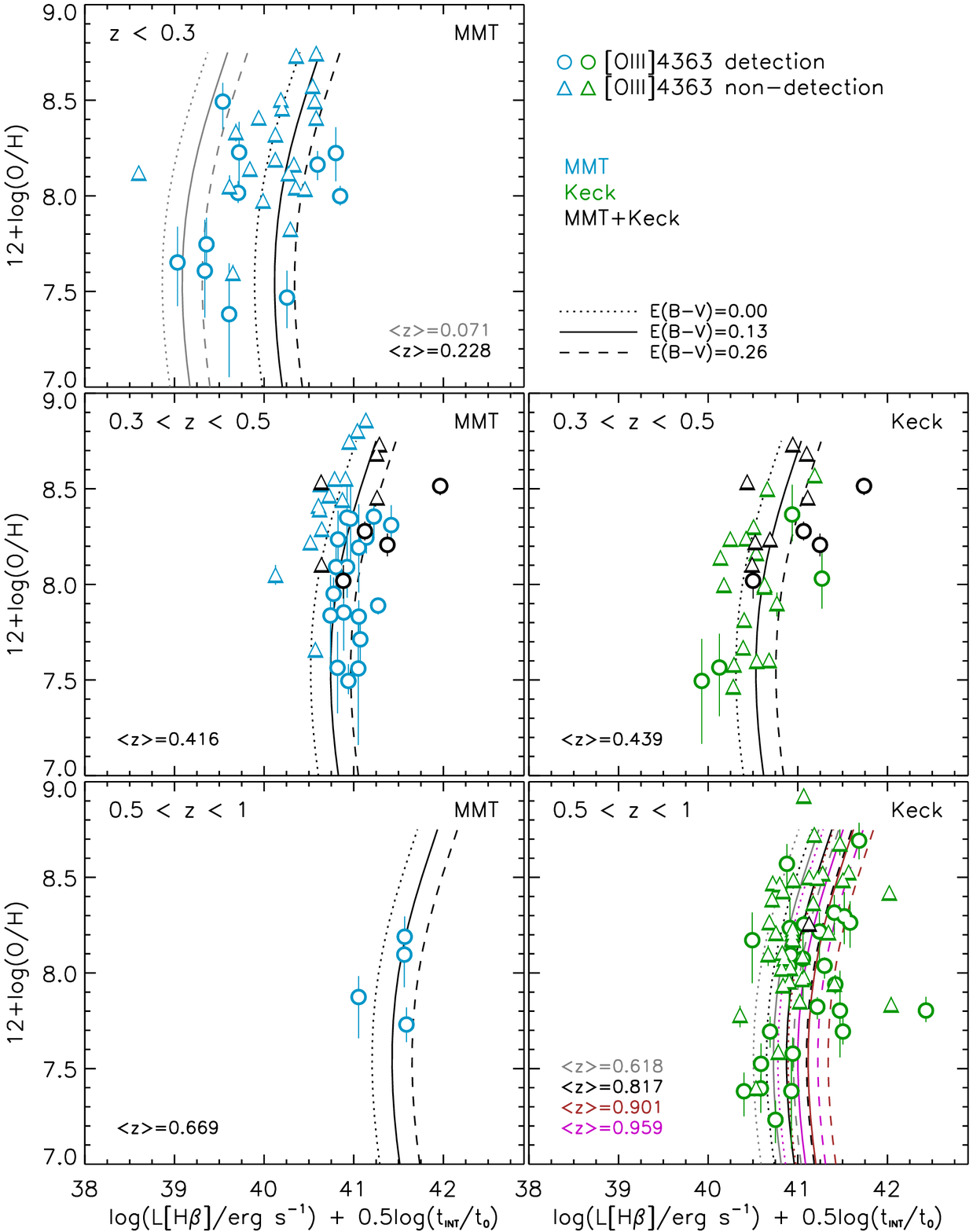}
  \caption{Oxygen abundances as a function of observed \Hb\ luminosity to illustrate
    the selection function of our spectroscopic survey with MMT (left) and Keck (right).
    Our \OIIIa-detected (circles) and \OIIIa-non-detected (triangles) samples from
    MMT (light blue), Keck (green), and both (black) are overlaid. The dotted, solid,
    and dashed curves correspond to the S/N = 3 limit on \OIIIa\ for three dust
    extinction possibilities that span the dispersion seen in our \OIIIa-detected
    galaxies. We illustrate these curves at different average redshifts with the \OIIIa\
    sensitivity estimated directly from the rms in the continuum of our spectroscopic
    data. To account for the varying integration time ($t_{\rm int}$) for each individual
    source, we normalize sensitivity to $t_0$ = 120 minutes. See Section~\ref{sec:bias}
    for further details.}
  \label{fig:sens}
\end{figure*}

To assess the selection function of our study, we examine the detectability of
\OIIIa\ with MMT and Keck as a function of redshift, metallicity, \Hb\ luminosity
(i.e., SFR), and dust attenuation.
To determine the $\mathcal{R}$ line ratio, we adopt a relation between \Te\ and \OH\
that is empirically based on our sample of \OIIIa\ detections:
\begin{equation}
  t_3 = 28.767 - 5.865 x + 0.306x^2,
\end{equation}
where $t_3\equiv T_e$(\OIII)/10$^4$ K and $x\equiv \OHm$. This O/H--\Te\ relation
is similar to that of \cite{nic14}, which is based on several local samples (see
their Figure 2).\footnote{While our relation is offset by $\sim$0.1 dex toward a
  higher metallicity at a given \Te, we note that the difference is due to the
  assumed relation between \Te(\OII) and \Te(\OIII). Here we use the
  \citetalias{and13} relation. We find that adopting the \cite{izo06b}
  \Te(\OII)--\Te(\OIII) relation, which is similar to that of \cite{nic14}, would
  yield a O/H--\Te\ relation that agrees (within measurement uncertainties) with
  \cite{nic14}.}
Then we determine the \OIII$\,\lambda$5007/\Hb\ flux ratio as a function of
metallicity by adopting $\log({\rm O}^+/{\rm O}^{++})=-0.114$, and the following
equation from \cite{izo06b}:
\begin{eqnarray}
  \nonumber
  12+\log{\left(\frac{{\rm O}^{++}}{{\rm H}^+}\right)} =
  \log{\left[\frac{F(\OIII\,\lambda\lambda4959,5007)}{F({\rm H}\beta)}\right]} + \\
  6.200 + \frac{1.251}{t_3} - 0.55\log{t_3} - 0.014t_3.
  \label{eqn:O++}
\end{eqnarray}
\noindent
This value of $\log({\rm O}^+/{\rm O}^{++})$ is the average of our \OIIIa-detected
sample, which does not appear to be dependent on \Te\ across $10^4$--$2.5\times10^4$ K.
We also examine local galaxies from \cite{berg12} and find no evidence for a \Te\
dependence across $10^4$--$2\times10^4$ K with an average
$\log({\rm O}^+/{\rm O}^{++})=-0.166$ that is similar to our measured average. The
combination of $\mathcal{R}$, \OIII\,$\lambda$5007/\Hb\ flux ratio, \Hb\ luminosity,
and redshift determines the \OIIIa\ line flux:
\begin{equation}
  F(\OIII\,\lambda4363) = \mathcal{R} \frac{1.33F(\OIII\,\lambda5007)}{F(\Hbe)} \frac{L(\Hbe)}{4\pi d_L^2},
\end{equation}
where $d_L$ is the luminosity distance and the \OIII\ $\lambda$5007/$\lambda$4959
flux ratio is $\approx$3 \citep{sto00}. We illustrate in Figure~\ref{fig:sens} the
average 3$\sigma$ \OIIIa\ line sensitivity for the MMT and Keck spectra. Here, the
sensitivity is computed by measuring the rms in the continuum of the spectra. We
illustrate the effects of dust attenuation on the expected \OIIIa\ line flux by
considering three different \EBVa\ values: 0.13 (the average in our \OIIIa\ sample),
0.0, and 0.26 (the range encompasses $\pm$1$\sigma$). The curves of \OIIIa\ line
sensitivity are computed from MMT (Keck) spectra for four (five) average redshifts
and overlaid in this figure with different colors. Because the on-source exposure
time varies by a factor of few to several, we have normalized all estimates to two
hours of integration ($t_0$). The observed points in Figure~\ref{fig:sens} account
for the individual integration times with an offset to the \Hb\ luminosity of
$0.5\log{(t_{\rm int}/t_0)}$.
Typically, it can be seen that our \OIIIa-detected galaxies lie to the right of our
line sensitivities, while the \OIIIa-non-detected galaxies lie to the left
of the line sensitivity.

Figure~\ref{fig:sens} demonstrates that the sensitivity to detect \OIIIa\ at solar
(half-solar) metallicity, assuming the same SFR or \Ha\ or \Hb\ luminosity, is on
average 2.9 (1.7) times lower than at \OH\ $\leq$ 8.0. This suggests that the high
stellar mass end (above 10$^9$ \Msun) of our \MZ\ relations is likely biased toward
lower metallicity, and that the \MZ\ relations could be steeper than reported in
Section~\ref{sec:MZ}. We note, however, that this bias is relatively modest (less
than a factor of 3 in sensitivity); thus, stacking at least $\sim$25 MMT and Keck
spectra of metal-rich galaxies will yield average detections of \OIIIa\ that are
significant above S/N = 5 at $z\sim1$. A forthcoming paper of \Sname\ will explore
measurements from stacking, and further examine our selection bias.


\subsection{Comparison with Previous \Te-based Abundance Studies}
{\it Narrow-band-selected sample}. One of the first studies to use the narrow-band
imaging technique to select high-EW emission-line galaxies to obtain \Te-based
metallicity was \cite{kak07}. They targeted narrow-band-excess emitters in the GOODS
fields with Keck/DEIMOS and obtained 23 galaxies with $\ge$3$\sigma$ detection of
\OIIIa\ \citep{hu09}. While our \OIIIa-detected sample is similar to theirs,
\cite{hu09} mostly measured \Te\ below \OH\ $\sim$ 8.0, whereas our sample spans a
wider range in metallicity at a given stellar mass or $M_B$.

{\it Magnitude-limited sample}. Our SDF \OIIIa\ sample at $z=0.2$--1 overlaps closely
in the \MZ\ plane with those measured by DEEP2 at $z\sim0.8$
\citep[see Figure~\ref{fig:MZ}]{ly15}. The main apparent difference is that the DEEP2
galaxies are from a magnitude-limited (i.e., \Mstar-limited) sample, which selects
galaxies above \Mstar\ $\sim10^{8.5}$ \Msun, and therefore higher metallicity.

In contrast, the \cite{amo14,amo15} \OIIIa-detected samples from VUDS and zCOSMOS,
respectively, are strongly biased to only metal-poor galaxies at all stellar masses.
This is well-illustrated in Figure~\ref{fig:MZ} where nearly all of their galaxies
are below the median of our galaxies at $z=0.5$--1 (solid brown line in the lower
left panel). These surveys obtain spectra at a lower resolution ($R\sim200$), which
limits the sensitivity to the detection of weak emission lines, particularly detecting
\OIIIa\ in more metal-rich galaxies.
This strong selection explains why galaxies from the \cite{amo14,amo15} samples have
systematically lower metallicities than galaxies in other samples.

The \cite{izo14} and \cite{berg12} samples at low redshift also show systematically
lower O/H than our SDF galaxies at $z\lesssim0.3$. \cite{izo14} selected galaxies
from SDSS with high \Hb\ EWs and at $z\sim0.2$. Because SDSS is a shallow
magnitude-limited survey, their sample selection biased them toward more massive
galaxies (\Mstar\ $\gtrsim10^{9}$ \Msun) with lower metallicities. \cite{berg12}
also reported that their sample has lower abundances at a given stellar mass
when compared with other local \MZ\ studies \citep{lee06}. It appears that our
$z\lesssim0.3$ \MZ\ relation is consistent with the \MZ\ relation of \cite{lee06};
however, we find a steeper slope.

\subsection{Comparison with Predictions from Galaxy Formation Simulations}

As discussed earlier, the shape and evolution of the \MZ\ relation are important
constraints for galaxy formation models, because the heavy-element abundances are set
by enrichment from star formation, with dilution and loss from gas inflows and
outflows, respectively \citep[see][and references therein]{som15a}. 
Efforts have been made to predict the \MZ\ relation from large cosmological
simulations that either hydrodynamically model the baryons in galaxies
\citep[e.g.,][]{dave11,obr14,vog14,sch15,ma16} or adopt semi-analytical models with
simple prescriptions for the baryonic physics \citep{gon14,lu14,por14,hen15,cro16}.

\begin{figure*}
  \epsscale{1.1}
  \plottwo{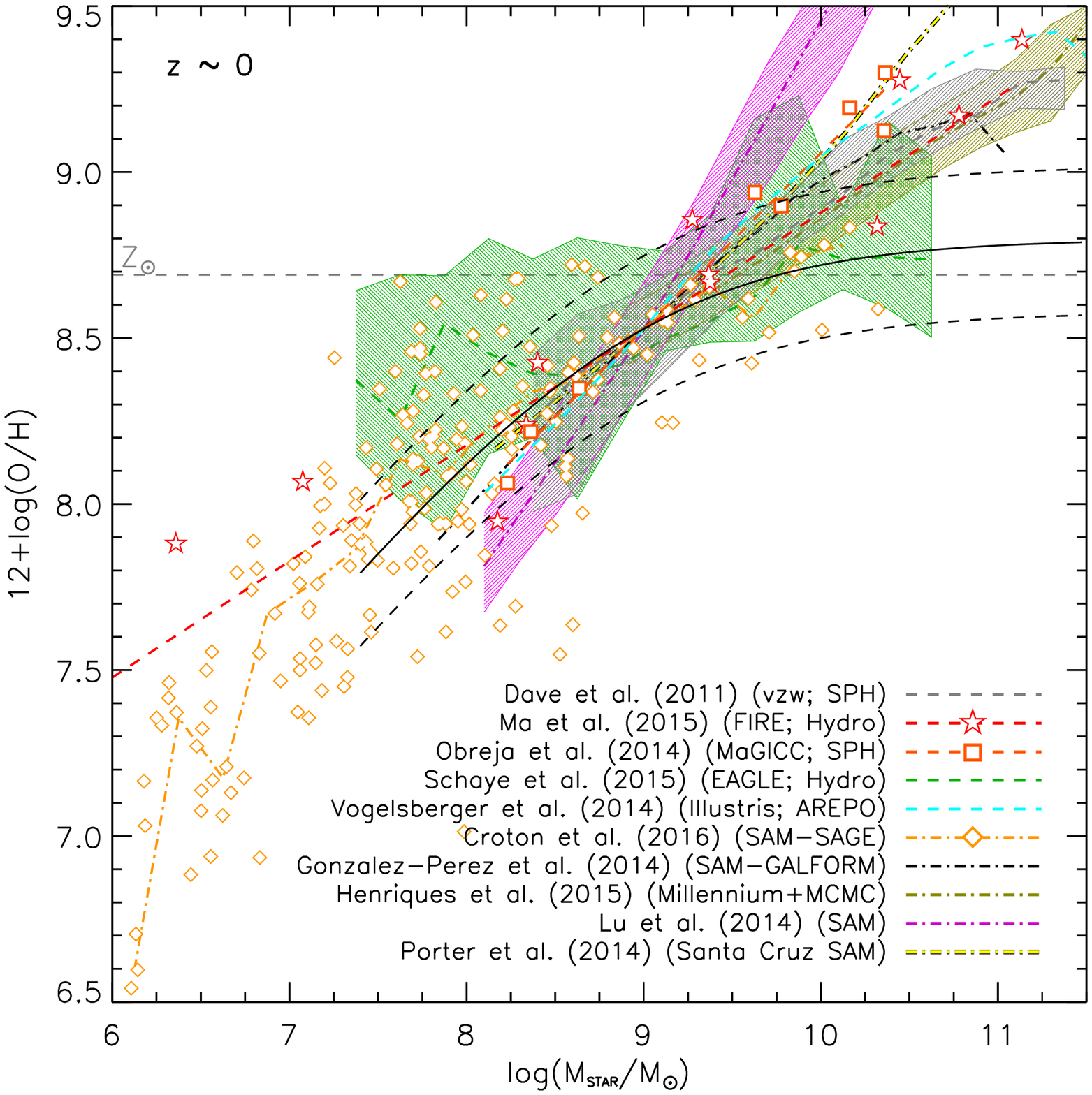}{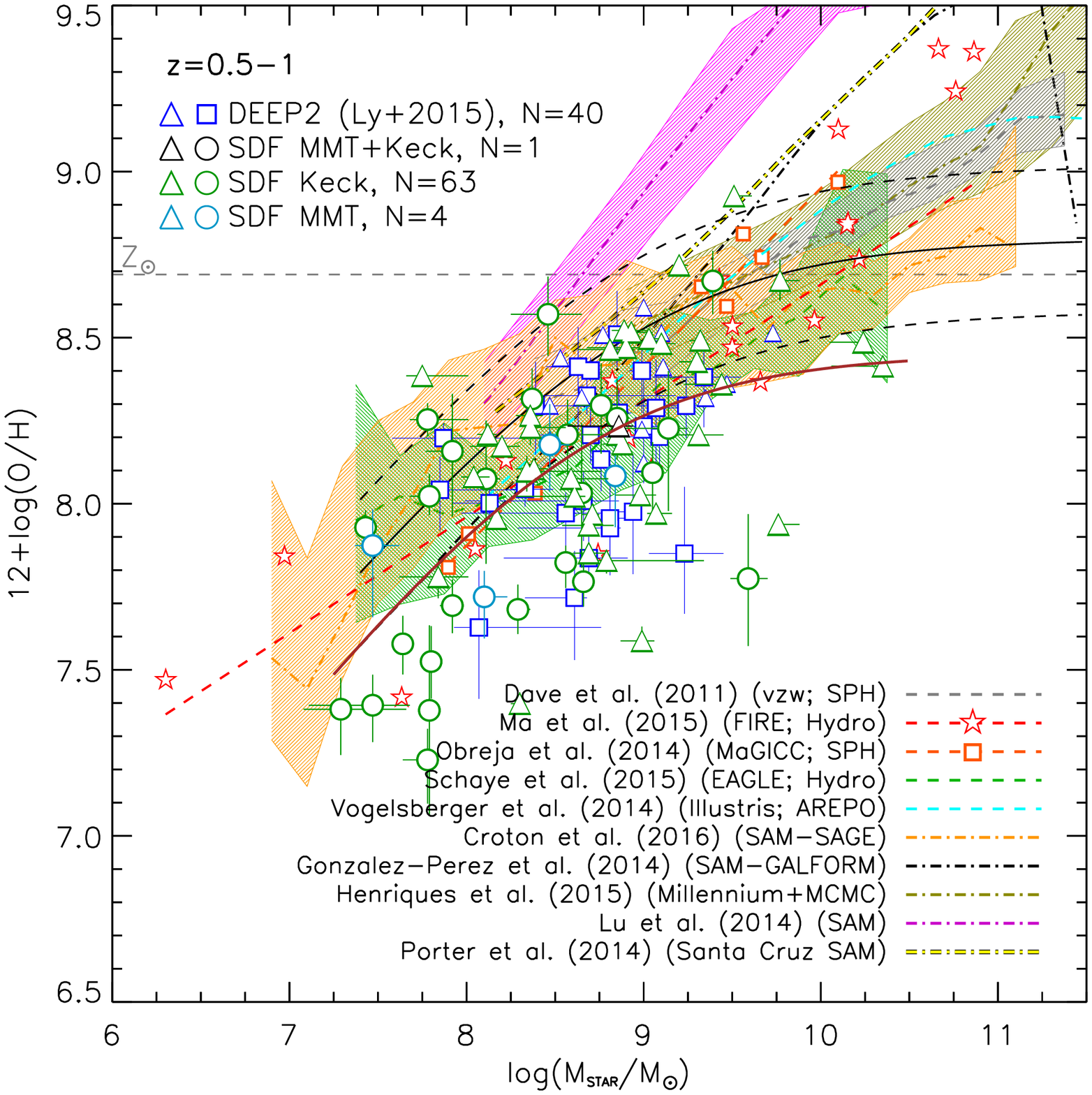}
  \caption{Comparison of the \Te-based \MZ\ relation at $z\sim0$ (left) and
    $z=0.5$--1 (right; lower left panel in Figure~\ref{fig:MZ}) against the $z\sim0$
    and $z\sim1$ predictions from galaxy formation numerical simulations. The light
    blue, green, and black symbols are galaxies from the SDF MMT, Keck, and MMT+Keck
    samples, respectively. Circles (triangles) illustrate galaxies with \OIIIa\
    detections (upper limits). The DEEP2 sample \citep{ly15} with \OIIIa\ detections
    (upper limits) is overlaid as dark blue squares (triangles). The best fit to the
    SDF and DEEP2 galaxies, which is shown in Figure~\ref{fig:MZ}, is also overlaid
    as the brown solid curve. Overlaid by the black solid and dashed lines is
    the local \MZ\ relation \citepalias{and13}. Hydrodynamical simulations are
    shown for \citet[gray dashed lines and gray shaded regions encompassing the
      16th and 84th percentile]{dave11}, \citet[FIRE; red dashed lines with
      individual galaxies shown as red stars]{ma16}, \citet[MaGICC; red--orange dashed
      lines with individual galaxies shown as red--orange squares]{obr14},
    \citet[EAGLE; green dashed lines and green shaded regions encompassing the 16th
      and 84th percentile]{sch15}, and \citet[Illustris; cyan dashed lines]{vog14}.
    In addition, we overlay semi-analytical predictions from \citet[orange
      dotted--dashed lines with orange diamonds on the left and orange shaded regions
      for $1\sigma$ dispersion]{cro16},
    \citet[GALFORM; black dotted--dashed lines]{gon14}, \citet[olive dotted--dashed
      lines and olive shaded regions]{hen15}, \citet[purple dotted--dashed lines and
      purple shaded regions encompassing the 16th and 84th percentile]{lu14}, and
    \citet[Santa Cruz; yellow dotted--dashed lines with black outlines]{por14}. For
    the EAGLE simulation, we limit predictions to galaxies above \Mstar\
    $= 2.3\times10^7$ \Msun, which corresponds to 10$^2$ star particles.
    \cite{sch15} caution against using metallicity predictions below 10$^4$ particles
    where resolution limits can over predict metallicity. We normalize the oxygen
    abundances of all theoretical/numerical predictions at $M_{\star} = 10^9$ \Msun\
    for $z\sim0$ \citepalias{and13}.}
  \label{fig:MZ_Theory}
\end{figure*}

We examine how well these numerical models and simulations predict the \MZ\ relation
in Figure~\ref{fig:MZ_Theory}. Here we compare $z\sim0$ predictions against
\citetalias{and13} in the left panel, and compare $z\sim1$ predictions against our
sample of $z=0.5$--1 galaxies in the right panel. We note that the normalization of
these predictions for the \MZ\ relation is dependent on the nucleosynthesis yield,
which is only accurately measured to $\sim$50\% (R. Dav{\'e} and K. Finlator 2015,
private communication). Thus, the normalization cannot be used to compare with
observations. For this reason, we normalize all \MZ\ relation predictions to
\OH\ = 8.5 at $M_{\star} = 10^9$ \Msun\ (at $z\sim0$; consistent with
\citetalias{and13}), and examine relative evolution. For simplicity and consistency,
predictions from hydrodynamic models are indicated by the dashed lines while
semi-analytical model predictions are denoted by the dotted--dashed lines.

First, we consider the predictions from the \textit{vzw} simulation by \cite{dave11},
which adopts ``momentum-conserving'' stellar winds. Their result is illustrated in
Figure~\ref{fig:MZ_Theory} by the gray dashed lines with the gray shaded regions
encompassing the 16th and 84th percentile.
At \Mstar\ $\sim 10^9$ \Msun, the slope in their \MZ\ relation is consistent with what
we and \citetalias{and13} measure. However, there are two issues with their predictions:
(1) the decline in abundances with redshift (from $z=0$ to $z\sim1$) that they measure
\citep[$\sim$0.1 dex; see Figure 2 in][]{dave11} is much lower than what we observe
($\approx$0.25 dex). (2) They predict a steep \MZ\ relation at higher stellar masses
at all redshifts. This was not seen in our observations or by \citetalias{and13}.
Unfortunately, the models from \cite{dave11} are unable to probe galaxies below
\Mstar\ $\approx10^{8.4}$ \Msun, where a steepening of the \MZ\ slope is seen at
$z\sim0$ and $z\sim1$.

Next, we compare our results against ``zoom-in'' hydrodynamical simulations from the
MaGICC \citep[Making Galaxies in a Cosmological Context;][]{bro12} and FIRE
\citep[Feedback in Realistic Environments;][]{hop14} projects. While these simulations
consist of much fewer galaxies than \cite{dave11}, they provide higher spatial
resolution on individual galaxies to resolve the structure of the ISM, star formation,
and feedback, and span a wider range in galaxy stellar masses. For MaGICC, the results
from \cite{obr14} are illustrated by the red--orange squares with the best linear fit
shown by the red--orange dashed lines. For FIRE, the redshift-dependent linear function
described in \cite{ma16} is illustrated by the red dashed lines in
Figure~\ref{fig:MZ_Theory} with red stars for individual
galaxies.\footnote{The metallicity normalization results in a small offset of 0.03 dex.}
Relative to \citetalias{and13}, \cite{obr14} measure a slightly steeper slope at 
\Mstar\ $\sim10^9$ \Msun\ than that found at \cite{dave11}, although this is within
the uncertainties. At $z\sim0.7$, \cite{obr14} find that abundances are lower by
$\approx$0.2 dex than at $z\sim0$, which is roughly consistent with our observed \MZ\
relation evolution.
Similar to \cite{dave11}, \cite{ma16} measures a slope that is consistent with our
sample and the results from \citetalias{and13} at \Mstar\ $\sim10^9$ \Msun. However,
the FIRE simulations find that abundances are lower by $\approx$0.25 dex at $z\sim0.8$
than at $z\sim0$, which is consistent with our results. Because of the limited sample
size of the MaGICC and FIRE simulations (only 7 and 24 galaxies at $z\sim1$,
respectively), constraining the shape of the \MZ\ relation is difficult. Furthermore,
the MaGICC (FIRE) simulations only have two (three) $z\sim1$ dwarf galaxies below
\Mstar\ $\sim10^8$ \Msun. These galaxies can provide the strongest constraints on
stellar winds from the \MZ\ relation. 

In addition, we also consider the predictions of \cite{vog14} from the Illustris
simulations,\footnote{The data set can be found at: \url{http://www.mit.edu/~ptorrey/data.html}.}
which are overlaid in Figure~\ref{fig:MZ_Theory} as the cyan dashed lines. Similar to
\cite{dave11}, \cite{obr14}, and \cite{ma16}, they predict an \MZ\ slope that is
consistent with \citetalias{and13} and our $z\sim1$ sample; however, much like
\cite{dave11}, the amount of evolution predicted at \Mstar\ $\sim10^9$ \Msun\ is
inconsistent with the $\approx0.25$ dex evolution that we observe. Also, the steep
\MZ\ slope at the high mass end that \cite{vog14} predict is inconsistent with our
observations and those of \citetalias{and13}.

Finally, we also overlay the predictions from the EAGLE hydrodynamical simulations
\citep{cra15,sch15} in Figure~\ref{fig:MZ_Theory} as green dashed lines and green
shaded regions encompassing the 16th and 84th percentile. The predictions are
obtained from the public catalog
\citep{mca16}.\footnote{\url{http://www.eaglesim.org/database.php.}
  We use the simulation with the highest particle resolution, Recal-L025N0752, and
  require SFR $>$ 0 for two snapshots, $z=1$ and $z=0$. Different sets of EAGLE
  simulations yield different results for the \MZ\ relation
  \citep[see Figure 13 of][]{sch15}; Recal-L025N0752 provides the best agreement
  with the \MZ\ relation.} The EAGLE simulation results agree with the observed
shape of the \MZ\ relation at $M_{\star} \gtrsim 3\times10^8$ \Msun\ for $z\sim0$
\citepalias{and13} and at $M_{\star} \gtrsim 3\times10^8$ \Msun\ for $z\sim1$. At
lower stellar mass it predicts a shallow \MZ\ slope, which is inconsistent with
observational results at $z\sim0$. However, this shallow slope is believed to be
caused by poor resolution because the turnover occurs when the number of star
particles falls below $\sim10^4$ \citep{sch15}. The EAGLE simulation does predict
$\approx$0.2 dex evolution in the \MZ\ normalization, which is consistent with
our observed evolution.

For completeness, we also overlay the predictions from several semi-analytical
models. These models adopt different assumptions and prescriptions. The predictions
for \cite{cro16}\footnote{Their latest results are obtained from the Theoretical
  Astrophysical Observatory (\url{https://tao.asvo.org.au/tao/}) using the largest
  simulated area, ``COSMOS''.} are shown by the dotted--dashed orange line and orange
diamonds\footnote{The sample size is small, so individual galaxies are shown rather
  than including a shaded region for a poorly measured dispersion.} for $z\sim0$ and
the dotted--dashed orange line and orange shaded region for $z\sim1$. We also overlay
\cite{gon14} as a black dotted--dashed line, \cite{hen15} as the olive dotted--dashed
line with olive shaded region indicating the 16th and 84th percentile, \cite{lu14}
as the purple dotted--dashed line with purple shaded region indicating 68\% dispersion, 
and \cite{por14} as the yellow dotted--dashed line with black outlines.

We note that none of these semi-analytical models predict a moderate ($\approx$0.25
dex) evolution in the \MZ\ relation at $z\lesssim1$. They either find no evolution or
no more than 0.1 dex. \cite{cro16}, \cite{hen15}, and \cite{por14} do predict a \MZ\
slope at \Mstar\ $\sim10^9$ \Msun\ that is consistent with observations at $z\sim0$
and $z\sim1$. The semi-analytical model that disagrees significantly from observations
is \cite{lu14}. They predict the steepest \MZ\ slope and higher metallicities at a
given stellar mass for $z\sim1$. \cite{cro16} is able to reproduce the shape of the
\MZ\ relation at $z\sim0$; however this is not a surprise because the local \MZ\
relation \citep{tre04} is used as a secondary constraint in their model.

Given these comparisons, we find that the only models that can reproduce both the
observed evolution in the \MZ\ relation ($\approx$0.25 dex) and the slope at \Mstar\
$\sim 10^9$ \Msun\ are the FIRE and EAGLE simulations. As discussed above, the FIRE
simulations provide the highest spatial resolution to resolve stellar feedback
\citep{hop14}, and the EAGLE simulation with the best agreement with the local \MZ\
relation has the highest particle resolution. While these results suggest that
resolving the physical processes in the ISM is critical for further understanding
the chemical enrichment process in galaxies, there has been some success in yielding
the same \MZ\ relation at multiple epochs with different resolutions \citep{tor14}.
This would suggest that what may be more important for lower resolution simulations
is correctly handling the baryonic gas flows on subresolution scales.

One reason why previous numerical simulations have not used the \MZ\ relation as an
observational constraint is the growing concern that ``strong-line'' metallicity
diagnostics may not be valid for higher redshift galaxies
\citep[e.g.,][]{ste14,san15,cow16,dop16}. However, now that the evolution of the
temperature-based \MZ\ relation has been measured, we encourage forthcoming models
and simulations to utilize the evolution of the \Te-based \MZ\ relation as an
important constraint for galaxy formation models. In addition, we encourage future
work to probe galaxies below \Mstar\ $\sim 10^8$ \Msun, because this remains an
unexplored parameter space where observations find a steep \MZ\ dependence (see
Figure~\ref{fig:MZ_Theory}). Finally, should computing infrastructures allow,
improving the particle resolution of large-scale galaxy simulations and using the
``zoom-in'' technique for more detailed studies are additional improvements that
may provide a better understanding of the gas flows in galaxies.


\section{CONCLUSIONS}
\label{sec:End}

We have conducted an extensive spectroscopic survey of $\approx$1900 emission-line
galaxies in the SDF with MMT/Hectospec and Keck/DEIMOS. Our spectroscopy detected
\OIIIa\ in \Ndet\ galaxies and provided robust \OIIIa\ upper limits for \Nrel\
galaxies. These measurements provide us with oxygen abundances from measuring the
electron temperature (\Te), and enable the first systematic study of the evolution
of the \MZ\ relation to $z\sim1$ using only the \Te\ method. We find that the \MZ\
relation evolves toward lower metallicity at fixed stellar mass proportional to
$(1+z)^{\zslope}$. In addition, we are able to measure the shape of the \MZ\ relation
at $z\approx0.5$--1. The shape is consistent with the local relation determined by
\citetalias{and13}, indicating a steep slope at the low-mass end, a flattening in
metallicity at $M_{\star} \sim 10^9$ \Msun, and abundances that are lower by
$\approx$0.25 dex at all stellar masses. We also examine whether the \MZ\ relation
has a secondary dependence on SFR such that galaxies with higher sSFR have reduced
metallicity. Our sample suggests that the SFR dependence is mild, and is at most
only a sixth as strong as that seen in local galaxies \citepalias{and13}. The weak
dependence on SFR may be due to large dispersion ($\approx0.3$ dex) that cannot be
attributed to measurement uncertainties, and a selection against metal-rich galaxies
with low SFR.
For the latter, we examine the selection function as a function of metallicity,
SFR, dust reddening, and redshift. We find that we mitigate the selection bias by
including a substantially large sample of reliable non-detections that have lower
sSFR by 0.5 dex over a wide range in stellar mass.

We also compare our \MZ\ relation results against predictions from semi-analytical
and hydrodynamic galaxy formation models. Specifically, we find good agreement on
the slope of the \MZ\ relation and its evolution with ``zoom-in'' simulations from
FIRE \citep{ma16} and high-resolution cosmological simulations from EAGLE \citep{sch15}.

Based on our analyses between observations from \Sname\ and galaxy formation
simulations, we suggest the following courses of action for forthcoming theoretical
studies on chemical enrichment: (1) utilize the evolution of the \Te-based \MZ\
relation as an important constraint for galaxy formation models, (2) simulate galaxies
below \Mstar\ $\sim10^{8}$ \Msun\ where observations suggest a steep \MZ\ relation
at both $z\sim0$ and $z\sim1$, (3) improve the particle resolution of large-scale
galaxy formation simulations, and (4) further use the ``zoom-in'' technique for
detailed examination of the ISM (e.g., resolving stellar feedback processes). These
improvements, combined with observational data of low-mass galaxies, will facilitate
a better physical understanding of the baryonic processes occurring within galaxies.

Our \OIIIa-detected sample includes a large number of extremely metal-poor galaxies
($\OHm\leq7.69$ or $\leq$0.1\zsun); it is the largest sample of extremely metal-poor
galaxies at $z\gtrsim0.2$. We argue that local surveys (e.g., SDSS) have not
identified many extremely metal-poor galaxies because they are magnitude-limited and
generally miss galaxies below \Mstar\ $\sim 10^8$ \Msun. Emission-line surveys that
utilize narrow-band imaging or grism spectroscopy are able to increase the efficiency
of identifying extremely metal-poor galaxies by detecting the nebular emission. Our
most metal-poor galaxy, Keck06, has an oxygen abundance that is similar to I Zw 18.
We also find that our high-sSFR galaxies are similar to typical $z\sim2$ galaxies in
terms of gas-phase metallicity and ionization parameter \citep{sha15}. This suggests
that a sample of analogs to $z\gtrsim2$ star-forming galaxies are available at
$z\lesssim1$ for more detailed spectroscopic studies.

\acknowledgements
We thank the anonymous referee for comments that improved the paper. The DEIMOS data
presented herein were obtained at the W.M. Keck Observatory, which is operated as a
scientific partnership among the California Institute of Technology, the University
of California, and the National Aeronautics and Space Administration (NASA). The
Observatory was made possible by the generous financial support of the W.M. Keck
Foundation.
The authors wish to recognize and acknowledge the very significant cultural role and
reverence that the summit of Mauna Kea has always had within the indigenous Hawaiian
community. We are most fortunate to have the opportunity to conduct observations from
this mountain.
Hectospec observations reported here were obtained at the MMT Observatory, a joint
facility of the Smithsonian Institution and the University of Arizona.
A subset of MMT telescope time was granted by NOAO, through the NSF-funded Telescope
System Instrumentation Program (TSIP). We gratefully acknowledge NASA's support for
construction, operation, and science analysis for the {\it GALEX} mission.
This research is supported by an appointment to the NASA Postdoctoral Program at the
Goddard Space Flight Center, administered by Oak Ridge Associated Universities and
Universities Space Research Association through contracts with NASA. CL is supported
by NASA Astrophysics Data Analysis Program grant NNH14ZDA001N. TN is supported by
JSPS KAKENHI Grant Number 25707010.
We thank Mithi de los Reyes for discussions that improve the paper.
We thank Darren Croton, Romeel Dav{\'e}, Violeta Gonzalez-Perez, Bruno Henriques,
Yu Lu, Xiangcheng Ma, Joop Schaye, Rachel Somerville, and Paul Torrey for
providing their theoretical data sets for comparison purposes and for discussions
that improved the paper.
We thank Alice Shapley for providing the MOSDEF data set for comparison purposes.
This paper utilizes the services of the Theoretical Astrophysical Observatory, which
is part of the All-Sky Virtual Observatory (ASVO) and is funded and supported by
Astronomy Australia Limited, Swinburne University of Technology, and the Australian
Government. The latter is provided though the Commonwealth's Education Investment Fund
and National Collaborative Research Infrastructure Strategy, particularly the National
eResearch Collaboration Tools and Resources (NeCTAR) Project.
We acknowledge the Virgo Consortium for making their simulation data available. The
EAGLE simulations were performed using the DiRAC-2 facility at Durham, managed by
the ICC, and the PRACE facility Curie based in France at TGCC, CEA,
Bruy{\'e}res-le-Ch{\^a}tel.
The Illustris simulation was run on the CURIE supercomputer at CEA/France as part of
PRACE project RA0844, and the SuperMUC computer at the Leibniz Computing Centre,
Germany, as part of project pr85je. Further simulations were run on the Harvard
Odyssey and CfA/ITC clusters, the Ranger and Stampede supercomputers at the Texas
Advanced Computing Center through XSEDE, and the Kraken supercomputer at Oak Ridge
National Laboratory through XSEDE.

{\it Facilities:} \facility{Subaru (Suprime-Cam)}, \facility{MMT (Hectospec)},
\facility{Keck:II (DEIMOS)}, \facility{{\it GALEX}},
\facility{Mayall (MOSAIC, NEWFIRM)}, \facility{UKIRT (WFCAM)}

\end{document}

%% file: tab1.final.tex
\begin{deluxetable}{crcccc}
  \tabletypesize{\scriptsize}
  \tablewidth{0pc}
  \tablecaption{Binned \MZ\ Relations}
  \tablehead{
    \colhead{$\log\left(M_{\star}/M_{\sun}\right)$}&
    \colhead{$N$}&
    \colhead{$<Z>$}&
    \colhead{Median $Z$}&
    \colhead{$\sigma_{\rm obs}$}&
    \colhead{$\sigma_{\rm int}$}\\
    \colhead{(dex)}&
    \colhead{}&
    \colhead{(dex)}&
    \colhead{(dex)}&
    \colhead{(dex)}&
    \colhead{(dex)}\\
    \colhead{(1)}&
    \colhead{(2)}&
    \colhead{(3)}&
    \colhead{(4)}&
    \colhead{(5)}&
    \colhead{(6)}}
  \startdata
  \multicolumn{6}{c}{$z\leq0.3$ (\Sname\ Only)}\\
  \pa7.25$\pm$0.25 &  8 & 7.85$^{+0.12}_{-0.12}$ & 7.83$^{+0.34}_{-0.09}$ & 0.35 & 0.30\\[1.5mm]
  \pa7.75$\pm$0.25 &  6 & 8.16$^{+0.07}_{-0.09}$ & 8.12$^{+0.23}_{-0.06}$ & 0.21 & 0.19\\[1.5mm]
  \pa8.25$\pm$0.25 &  3 & 8.53$^{+0.10}_{-0.11}$ & 8.46$^{+0.00}_{-0.05}$ & 0.17 & 0.17\\[1.5mm]
  \pa8.75$\pm$0.25 &  7 & 8.26$^{+0.12}_{-0.14}$ & 8.22$^{+0.18}_{-0.05}$ & 0.37 & 0.36\\[1.5mm]\hline
  \multicolumn{6}{c}{$0.3 < z\leq0.5$ (\Sname\ Only)}\\
  \pa7.25$\pm$0.25 &  5 & 7.90$^{+0.13}_{-0.10}$ & 7.95$^{+0.06}_{-0.11}$ & 0.27 & 0.25\\[1.5mm]
  \pa7.75$\pm$0.25 &  6 & 7.90$^{+0.12}_{-0.15}$ & 8.02$^{+0.22}_{-0.28}$ & 0.36 & 0.30\\[1.5mm]
  \pa8.25$\pm$0.25 & 12 & 8.14$^{+0.09}_{-0.08}$ & 8.24$^{+0.00}_{-0.14}$ & 0.36 & 0.34\\[1.5mm]
  \pa8.75$\pm$0.25 & 19 & 8.27$^{+0.06}_{-0.07}$ & 8.30$^{+0.05}_{-0.07}$ & 0.30 & 0.29\\[1.5mm]
  \pa9.25$\pm$0.25 & 10 & 8.15$^{+0.13}_{-0.12}$ & 8.19$^{+0.16}_{-0.19}$ & 0.41 & 0.39\\[1.5mm]\hline
  \multicolumn{6}{c}{$0.5 < z\leq1.0$ (\Sname\ + \citealt{ly15})}\\
  \pa7.50$\pm$0.25 &  5 & 7.63$^{+0.11}_{-0.14}$ & 7.58$^{+0.00}_{-0.20}$ & 0.26 & 0.22\\[1.5mm]
  \pa8.00$\pm$0.25 & 19 & 7.92$^{+0.07}_{-0.07}$ & 8.04$^{+0.03}_{-0.07}$ & 0.32 & 0.29\\[1.5mm]
  \pa8.50$\pm$0.25 & 32 & 8.10$^{+0.04}_{-0.05}$ & 8.10$^{+0.09}_{-0.06}$ & 0.25 & 0.23\\[1.5mm]
  \pa9.00$\pm$0.25 & 37 & 8.25$^{+0.04}_{-0.04}$ & 8.26$^{+0.04}_{-0.02}$ & 0.25 & 0.22\\[1.5mm]
  \pa9.50$\pm$0.25 & 11 & 8.41$^{+0.09}_{-0.08}$ & 8.38$^{+0.05}_{-0.03}$ & 0.28 & 0.26\\[1.5mm]
    10.00$\pm$0.25 &  3 & 8.37$^{+0.23}_{-0.16}$ & 8.49$^{+0.16}_{-0.00}$ & 0.38 & 0.38\\[1.5mm]
  \vspace{-3mm}
  \enddata
  \label{tab:binned_mz}
  \tablecomments{(1): Stellar mass bin. (2): Number of galaxies in each stellar mass bin, $N$.
    (3): Average \OH. (4): Median \OH. (5): Observed dispersion in \OH\ in each stellar mass bin.
    (6): Intrinsic dispersion in \OH\ after accounting for the average \OH\ measurement uncertainty,
    $\sigma_{\rm int} = \sqrt{\sigma_{\rm obs}^2 - \left<\Delta{\rm O/H}\right>^2}$.
    Uncertainties for averages and medians are reported at the 16th and 84th percentile. These
    uncertainties are determined by statistical bootstrapping: random sampling with replacement,
    repeated 10,000 times.}
\end{deluxetable}

%% file: tab2.final.tex
\begin{deluxetable}{rccc}
  \tabletypesize{\scriptsize}
  \tablewidth{0pc}
  \tablecaption{Best Fit to Binned \MZ\ Relations}
  \tablehead{
    \colhead{Redshift}&
    \colhead{\OH$_{\rm asm}$}&
    \colhead{$\log(M_{\rm TO}/M_{\sun})$}&
    \colhead{$\gamma$}\\
    \colhead{}&
    \colhead{(dex)}&
    \colhead{(dex)}&
    \colhead{}\\
    \colhead{(1)}&
    \colhead{(2)}&
    \colhead{(3)}&
    \colhead{(4)}}
  \startdata
        $z\sim0.1$\TA &  8.798                & 8.901                & 0.64\\[1.5mm]
        $z\leq0.3$    &  8.78$^{+0.11}_{-0.10}$ & 8.901\TB             & 0.47$^{+0.11}_{-0.11}$\\[1.5mm]
  $0.3 < z\leq0.5$    &  8.49$^{+0.07}_{-0.07}$ & 8.901\TB             & 0.29$^{+0.10}_{-0.09}$\\[1.5mm]
  $0.5 < z\leq1.0$    &  8.53$^{+0.06}_{-0.05}$ & 8.901\TB             & 0.57$^{+0.10}_{-0.08}$\\[1.5mm]
  $0.5 < z\leq1.0$    &  8.46$^{+0.34}_{-0.05}$ & 8.61$^{+0.57}_{-0.60}$ & 0.67$^{+0.30}_{-0.09}$\\[1.5mm]
  \vspace{-3mm}
  \enddata
  \label{tab:mz_best_fit}
  \tablecomments{(1): Redshift. (2): Asymptotic metallicity at the high stellar mass end
    of the \MZ\ relation. (3): Turnover mass in the \MZ\ relation. (4): Slope of the
    low-mass end of the \MZ\ relation. See Equation~(\ref{eqn:MZ}) and Section~\ref{sec:MZ}
    for further information. Uncertainties are reported at the 16th and 84th percentile, and
    are determined from the probability functions marginalized over the other two fitting
    parameters. Figure~\ref{fig:MZ_fit_contours} illustrates the confidence contours for all
    three fitting parameters.}
  \tablenotetext{1}{From \citetalias{and13}.}
  \tablenotetext{2}{The turnover mass was fixed to the value from \citetalias{and13}.}
\end{deluxetable}